\RequirePackage{fix-cm}
\documentclass[smallextended]{svjour3}       
\smartqed  
\usepackage{graphicx}

\usepackage{amsthm}
\usepackage{amssymb}
\usepackage{amsmath}
\usepackage{mathrsfs}
\usepackage{comment} 
\usepackage{algorithmic,algorithm}
\usepackage{shuffle}
\usepackage{graphicx} 
\usepackage{caption}
\usepackage{lipsum}
\usepackage{color}
\usepackage{enumerate}
\usepackage{subfigure} 
\usepackage{array}
\usepackage{enumerate}
\usepackage{setspace}
\usepackage{multirow}
\graphicspath{{\figs}}

\begin{document}

\title{Overview of Networked Control with Imperfect Communication Channels
}


\author{Yuting Zhu \and
        Liyong Lin \and
        Ruochen Tai \and
        Rong Su
}


\institute{Yuting Zhu \at
              \email{yuting002@e.ntu.edu.sg}           
           \and
           S. Author \at
              second address
}

\date{Received: date / Accepted: date}

\maketitle

\begin{abstract}
The paper focus on the networked supervisory control framework based on the discrete event systems with imperfect network, which can be divided into the centralized supervisory control and decentralized control. Then we reviewed the state-of-art networked control frameworks with observation channel delays and control channel delays separately, and the approach to compose the synthesized supervisor are constructed in untimed models and timed discrete event systems (TDES). The lossy property of communication channels is considered as well. 
\keywords{Networked control system \and Communication channel \and Supervisor Synthesis}
\end{abstract}

\section{Introduction}
\label{sec:network_control_intro}

Networked systems have been seen almost everywhere in our daily life today. They have been thoroughly studied in the systems and control community for about 20 years under the umbrella of multi-agent systems [...], leading to numerous publications that address a broad scope of topics such as cooperative/non-cooperative control, cyber physical system control, hybrid systems, distributed optimization, task planning, social sourcing and distributed learning, and cyber security analysis and control. One of the most important research aspects is how information is generated and propagated in the network, which directly affects the overall network performance. The discrete-event system community, in particular, the supervisory control community, has also been actively involved in this popular research trend, as illustrated in a large number of publications on modular control and decentralized/distributed control, where a target system is comprised of a set of local agents, interacting with each other via specific synchronization mechanisms, and each agent is managed by one or several local controllers via specific information fusion mechanisms. The goal is to ensure {\em safety}, i.e., no bad behaviours will happen, {\em liveness}, i.e., good behaviours will (eventually) happen, and {\em optimality}, i.e., the attained system performance should be the best among all possible ones.

There are two major control frameworks in the DES community, namely {\em decentralized control} and {\em modular control} - the latter may also called {\em distributed control}. The major difference between these two frameworks is whether each locally controlled subsystem $(G_i,S_i)$ is controllable and observable without assistance of other local supervisors, while jointly fulfilling the system requirements.

\begin{figure}[htb]
    \begin{center}
      \includegraphics[width=0.4\textwidth]{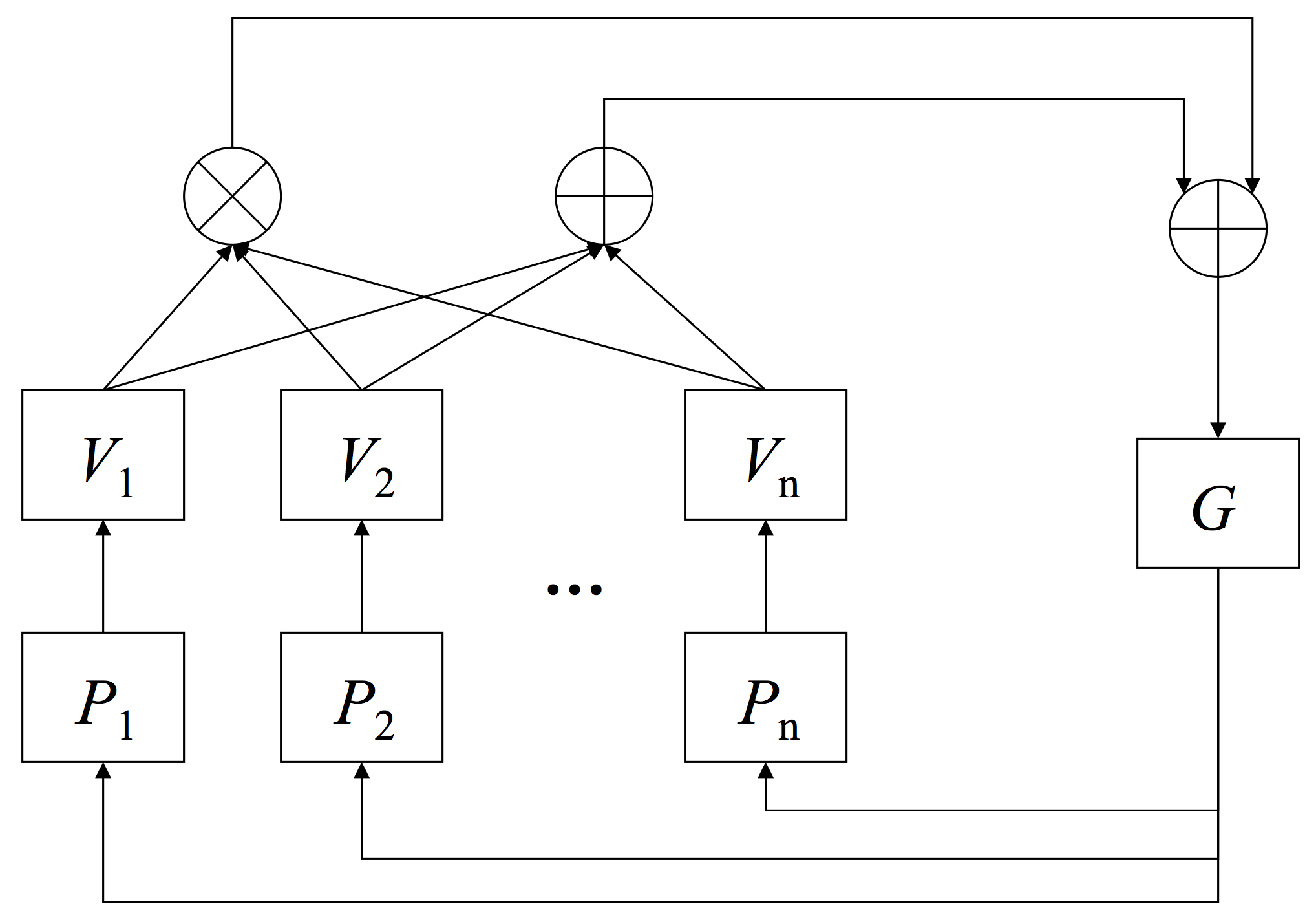}
    \end{center}
    \caption{A general architecture of decentralized control}
    \label{fig:FnT-Networked-Control-1}
\end{figure}

In a general decentralized supervisory control paradigm \cite{yoo2002general}, depicted in Figure \ref{fig:FnT-Networked-Control-1},
where the alphabet $\Sigma$ is covered by a set of local alphabets $\{\Sigma_i\subseteq\Sigma|i\in I=\{1,\cdots,n\}\}$, $\Sigma_{c,i}:=\Sigma_c\cap\Sigma_i=\Sigma_{c,e,i}\cup\Sigma_{c,d,i}$, and the plant $G$ is controlled by a set of local supervisors $V_i:P_{o,i}(L(G))\rightarrow \Gamma$ ($i\in I$) with $P_{o,i}:\Sigma^*\rightarrow (\Sigma_i\cap\Sigma_o)^*$ being the natural projection, via both the conjunctive fusion rule (denoted by ``$\oplus$'') and the disjunctive fusion rule (denoted by ``$\otimes$''). In the conjunctive fusion rule (CFR), for each $s\in L(G)$, we have

\begin{align}
        V_{c,i}(P_{o,i}(s)):=\{\sigma\in\Sigma_{c,e,i}\subseteq\Sigma_c\cap\Sigma_i|P_{o,i}^{-1}P_{o,i}(s)\sigma\cap\bar K  \neq\varnothing\}\cup (\Sigma_c-\Sigma_{c,e,i})\cup\Sigma_u.
\end{align}

The conjunctive permissive control map is $V_c:L(G)\rightarrow\Gamma$, where
\[(\forall s\in L(G))\, V_c(s):=\bigcap_{i\in I}V_{c,i}(P_{o,i}(s)).\]
In the disjunctive fusion rule (DFR), for each $s\in L(G)$, we have

    \begin{align}
    V_{d,i}(P_{o,i}(s)):=\{\sigma\in\Sigma_{c,d,i}\subseteq\Sigma_c\cap\Sigma_i|P_{o,i}^{-1}P_{o,i}(s)\sigma\subseteq\bar K\}  \cup \Sigma_u.
    \end{align}

The disjunctive anti-permissive control map is $V_d:L(G)\rightarrow\Gamma$, where
\[(\forall s\in L(G))\, V_d(s):=\bigcup_{i\in I}V_{d,i}(P_{o,i}(s)).\]
The general supervisory control map is $V_g:L(G)\rightarrow\Gamma$ where 
\[(\forall s\in L(G)) V_g(s):=V_c(s)\cup V_d(s).\]
To make this control architecture work, i.e., a given sublanguage $K\subseteq L_m(G)$ is equal to $L_m(V_g/G)$, the sublanguage $K$ must be {\em controllable}, {\em $L_m(G)$-closed}, and {\em co-observable}, which is defined as follows: 
\begin{itemize}
\item The language $K_c:=L(V_c/G)$ is {\em (C\& P) co-observable} w.r.t. $G$ and $(\Sigma_{c,e}:=\cup_{i\in I}\Sigma_{c,e,i},\{P_{o,i}|i\in I\})$, where for all $s\in\bar K$ and $\sigma\in\Sigma_{c,e}$,

\begin{align}
        s\sigma\in L(G)-\bar K_c\Rightarrow (\exists i\in I)P_{o,i}^{-1}P_{o,i}(s)\sigma\cap \bar K_c=\varnothing\wedge \sigma \in\Sigma_{c,e,i}.
\end{align}

\item The language $K_d:=L(V_d/G)$ is {\em (D\& A) co-observable} w.r.t. $G$ and $(\Sigma_{c,d}:=\cup_{i\in I}\Sigma_{c,d,i},\{P_{o,i}|i\in I\})$, where for all $s\in\bar K_d$ and $\sigma\in\Sigma_{c,d}$,
\[s\sigma\in \bar K_d\Rightarrow (\exists i\in I)P_{o,i}^{-1}P_{o,i}(s)\sigma\subseteq \bar K_d\wedge \sigma\in\Sigma_{c,d,i}.\]
\end{itemize}
Intuitively, (C\&P) co-observability ensures that each bad string can be identified by at least one local supervisor, and (D\&A) co-observability ensures that each good string can be confirmed by at least one local supervisor. The observability concepts are not compatible when more than one local supervisor exists, but both reduced to the same observability concept in the centralized framework.

Each local observation in the decentralized control strategy may be enhanced by allowing event communication among local supervisors. There are a lot of works on this topic focusing on the synchronous communication for control of decentralized DES, where the communication is assumed to involve zero delay. In \cite{barrett2000decentralized}, a novel information structure formalism is presented, which represents actions observable by each controller, which controllers communicate to other controllers, what symbols are communicated, when controllers initiate communication, and what information may be inferred by each of the controllers following any sequence of actions. Based on this structure, both anticipating controllers and myopic controllers are studied and characterized. \cite{ricker1999incorporating} \cite{ricker2008asymptotic} \cite{wang2008minimization} investigate the minimal communication policies for decentralized control where a communication policy is said to be minimal if removing one or more communications of event occurrences in the dynamic evolution of the system renders a correct solution incorrect. Specifically, the minimal communication problem is translated into one that can be solved on a Markovian mode in \cite{ricker2008asymptotic}, based on which the minimal cost communication protocol can be found by solving an optimization problem over a set of Markov chains. Under an assumption on the absence of cycles (other than self-loops) in the system model, \cite{wang2008minimization} proposes a polynomial time algorithm in the size of the state space of the plant for the synthesis of communication policies, which is an improvement compared with previous works. As a chapter devoted to the review of this research field, \cite{ricker2013overview} summarizes two approaches to the synthesis of communication protocols: state-based communication and event-occurrence communication. However, since these works involve zero delay, they might not be applied in some realistic scenarios where delays happen unavoidably in the shared communication network.

\begin{figure}
    \begin{center}
      \includegraphics[width=0.4\textwidth]{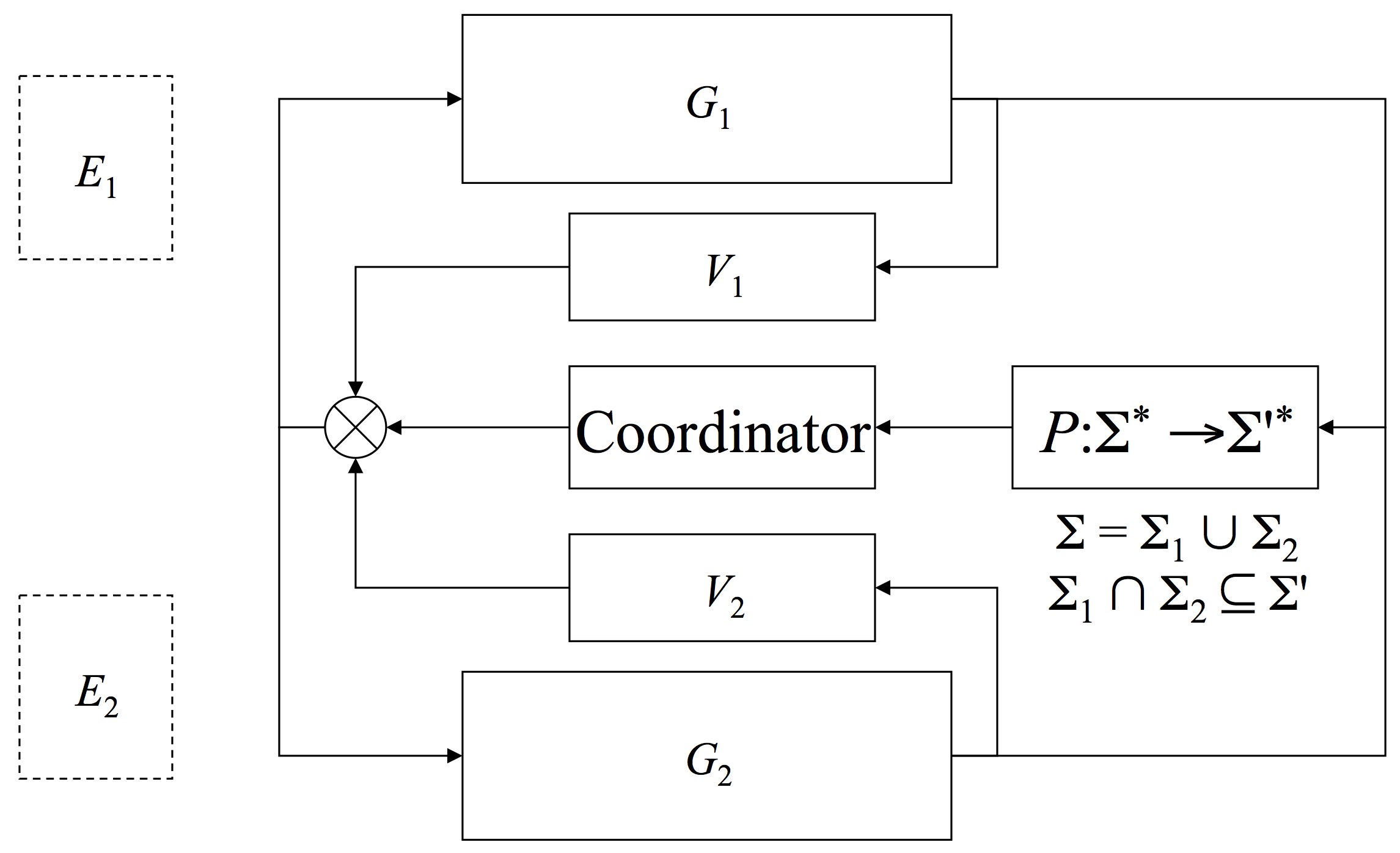}
    \end{center}
    \caption{A general architecture of modular control}
    \label{fig:FnT-Networked-Control-2}
\end{figure}

Notice that in the decentralized control framework, each local control law $V_i$ ($i\in I$) cannot ensure the local closed-loop behaviour $L(V_i/G_i)$ to be controllable and observable. Instead, it requires a genuine co-design of all local control laws $\{V_i|i\in I\}$ to ensure global controllability and observability with proper fusion rules accompanied by suitable concepts of (C\&P, D\&A) co-observability, i.e., for every single undesired string in the system, it requires a specific joint effort of all local supervisors to prevent the string from happening. Since typically we have an infinite number of undesired strings, it is usually undecidable \cite{tripakis2004decentralized} whether there exists a decentralized supervisor to achieve the  goal. To avoid this unpleasant undecidability issue faced by decentralized control, significant efforts have been made in developing a modular control (or distributed control) framework, where the set of all undesired behaviours is divided into a finite set of languages, each of which will be handled by one specific local supervisor. This will prevent us from enumerating each single undesired string, thus, lead to a terminable design procedure, at the price of reducing the solution space. The general architecture of modular control is illustrated in Figure \ref{fig:FnT-Networked-Control-2},
where there are multiple local components $\{G_i|i\in I\}$, where the alphabet of each $G_i$ is $\Sigma_i$. There are a finite number of local requirements $\{E_i\subseteq L_m(G_i)|i\in I\}$. The goal is to design a set of local control laws $\{V_i:L(G_i)\rightarrow \Gamma|i\in I\}$ together with a coordinator $C:P_C(L(||_{i\in I}G_i))\rightarrow \Gamma$, where $P_C:(\cup_{i\in I}\Sigma_i)^*\rightarrow \Sigma_C^*$ is the natural projection with $\Sigma_C\subseteq \cup_{i\in I}\Sigma_i$ being the alphabet of the coordinator, such that the following property holds: Let $G=||_{i\in I}G_i$.
\begin{itemize}
\item For each $i\in I$,
      \begin{itemize}
      \item $L_m(V_i/G_i)\subseteq E_i$;
      \item $L_m(V_i/G_i)$ is controllable and observable w.r.t. $G_i$ and $(\Sigma_{c,i},\Sigma_{o,i})$;
      \item $V_i/G_i$ is nonblocking.
      \item Supremality or maximality can be imposed on $L_m(V_i/G_i)$, depending on the choice of observability.
      \end{itemize} 
\item The closed-loop system $V/G$ is nonblocking, where $V=\wedge_{i\in I}V_i\wedge C$. 
\end{itemize}
Efficient synthesis methods have been developed in the literature based on either languages and bottom-up abstraction \cite{feng2008supervisory}, top-down decomposition \cite{komenda2015coordination}, or automata and bottom-up abstraction \cite{su2010model} \cite{su2010aggregative} \cite{su2011synthesis} to solve this problem, as the existence of each $V_i$ and $C$ is decidable.      

In contrast to existing works on multi-agent systems, where the quality of networked communication among agents, e.g., signal noises and disturbances, message delays and dropouts, plays one key role in system analysis and control, discrete-event system control theories rarely consider such imperfect communication, partially due to the modelling limitations - after all, most DES works rely on the regular language assumption. For example, in the standard Ramadge-Wonham supervisory control theory, it is assumed that event executions are {\em instantaneous} and {\em asynchronous}, which was later relaxed by introducing max-plus automata \cite{gaubert1995performance}, time-weighted automata \cite{su2011synthesis} and timed Petri nets, where events have durations and asynchrony of event executions refers to the starting times of relevant events, instead of their ending times. Yet, the key First-In-First-Out (FIFO) assumption of event generation and receiving must hold in each component and supervisor. Due to this assumption, details of network communication processes are not critically important, even when observable outputs of the plant may be delayed (but unanimously) or lost due to transmission failures. Details of relevant works on ``imperfect'' observationsdue to sensor failures are discussed in the Chapter on Fault Diagosis and Fault Tolerant Control. So one big question is: {\em what will happen if the FIFO assumption does not hold in the observation and command channels?}                

\begin{figure}[htb]
    \begin{center}
      \includegraphics[height=0.4\textwidth]{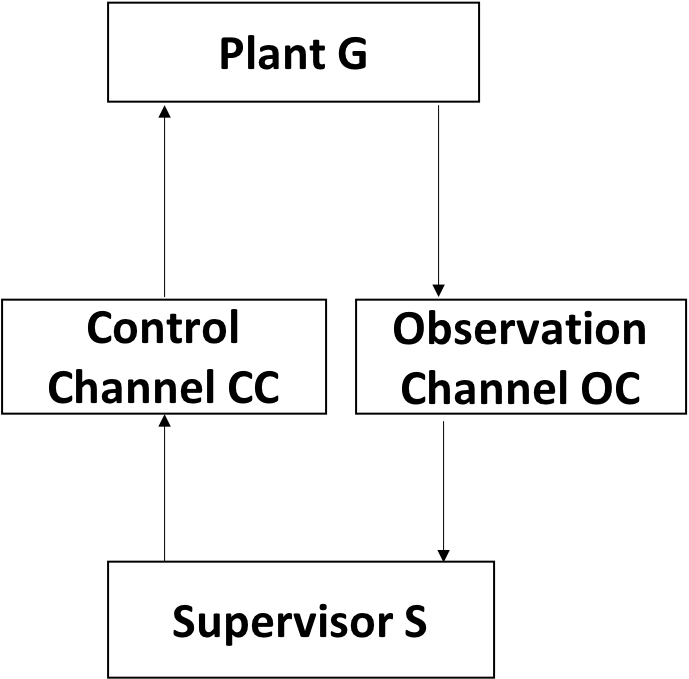}
    \end{center}
    \caption{A schematic diagram of the networked control systems}
    \label{fig:diagram}
\end{figure}

To facilitate a simple discussion, we first narrow ourselves to a simple network setup, depicted in Fig.~\ref{fig:diagram}, where there is one plant $G$, one supervisor $S$, one directed observation channel $OC$ from $G$ to $S$, and one directed control channel $CC$ from $S$ to $G$. 
Each observable output generated by the plant $G$ is transmitted to $S$ via $OC$, and each control command generated by $S$ is transmitted to $G$ via $CC$. In the Chapter of Cyber Security in Discrete Event Systems we have discussed models of deliberate attacks on $OC$ and $CC$, and relevant attack-resilient control techniques. Here, we will address the impact of channel imperfection on the overall performance of the closed-loop system $(G,S)$. More explicitly, we will only focus on message delays and dropouts in both $OC$ and $CC$. The first challenge is how to model channel delays imposed on observations and control commands, especially, delays that void the FIFO assumption. Due to computability and computational complexity concerns, current works mainly focus on regular languages or equivalent models. So it is a common practice to assume a known upper bound of delays for each message. When the message is not received by the upper bound, it is assumed lost or dropped out - so the upper bound also serves as a ``time-out'' mechanism. Each delay can be measured in time or in the number of events, depending on a user's needs. Once a delay model is specified, the next challenge is how to model interactions between delayed channels and the plant $G$ and supervisor $S$. This model is crucial, as it directly affects the subsequent concepts of {\em network controllability} and {\em network observability}, thus, determines whether a proposed networked control framework is practical feasible.\\         

After having proper channel delay models and channel-$(G,S)$ interaction models, the following fundamental questions need to be answered:
\begin{enumerate}
    \item How to model the  closed-loop system?
    \item What conditions may ensure the existence of a networked supervisor resilient to channel delays?
    \item How to synthesize such a network-delay resilient supervisor?
    \item Is it possible to carry out the synthesis  efficiently?
\end{enumerate}
In the next section, we shall review the state-of-art works for each different setup, aiming to answer the above questions. We will categorize existing frameworks based on specific models and synthesis strategies.   

\section{Review of state-of-art networked control frameworks}

\subsection{State-of-art networked control with $OC$ delays}
We organize existing publications based on whether they are related to only $OC$ delays or both $OC$ and $CC$ delays. It addresses synthesis, verification and computability issues.  
\subsubsection{$OC$ delays in decentralized control}
In \cite{park2007decentralized} the authors introduce a special $OC$ delay model, where, after each control command is sent, there may be several uncontrollable event firings in the system before a new control command is generated, i.e., message transmission in $OC$ is not instantaneous, and its maximum duration is determined by the maximum number of consecutive uncontrollable event firings. Three assumptions are made, which are listed below:
\begin{enumerate}
    \item Each locally controllable event is locally observable, i.e., $\Sigma_{c,i} \subseteq \Sigma_{o,i}$;
    \item The number of consecutive occurrences of uncontrollable events in the plant $G$ is upper bounded with a finite bound;
    \item Messages in both $OC$ and $CC$ (after fusion) are FIFO. 
\end{enumerate}
The authors introduce a new (C\&P) control law, extended from the standard architecture shown in Subsection \ref{sec:network_control_intro}, which is stated below: for each $s\in L(G)$,
\begin{center}$V_i(P_{o,i}(s)) := \{\sigma \in \Sigma_{c,i}|P_{o,i}^{-1}(P_{o,i}(s))\Sigma_u^*\{\sigma\}\cap\bar K\neq\varnothing\} \cup (\Sigma-\Sigma_{c,i}),
$\end{center}
where $\Sigma_K(s') = \{\sigma \in \Sigma|s'\sigma \in \bar K\}$. The basic idea is that, any event $\sigma$ after $s'$, observably identical to $s$, and a finite sequence of uncontrollable events $u$ is allowed by the supervisor $V_i$, as long as $s'u\sigma\in\bar K$. The final (C\&P) control law is shown as follows: for all $s\in L(G)$,
\[ V_{dec}(s) := \left\{
\begin{array}{ll}
\varnothing            &
(\exists t \in \Sigma^{*})(\exists u \in \Sigma_u^*) \\&s=tu\wedge (\forall u' \in \overline{\{u\}} -\{\varepsilon, u\}) \\ 
& S_{dec}(t) \neq \varnothing \wedge S_{dec}(tu') = \varnothing,\\
\cap_iV_i(P_{o,i}(s))       & \textrm{otherwise}.
\end{array} \right. \]
That is, the control law $V_{dec}$ will generate a new command, only when the current string $s$ is not a suffix string of $t$ via an uncontrollable sequence $u$, such that there is a control command at $t$, but there is no new command afterwards till now, due to observation delay in $OC$. The closed-loop behaviour is defined as follow: 
\begin{itemize}
\item $\epsilon \in L(V_{dec}/G)$, 
\item for all $s \in L(V_{dec}/G)$ and $\sigma \in \Sigma$ with $s\sigma \in L(G)$, $s\sigma\in L(V_{dec}/G)$ iff
$(\exists t \in \Sigma^*)(\exists u \in \Sigma_u^*)s=tu\wedge \sigma\in V_{dec}(t)\wedge (\forall v \in \overline{\{u\}}-\{\epsilon\}) V_{dec}(tv) = \varnothing.$
\end{itemize}
Similar to \cite{yoo2002general}, to make the proposed (C\&P) control law works, the authors extend the concept of (C\&P) co-observability to properly handle $OC$ delays.
\begin{definition}
\textnormal{A sublanguage $K \subseteq L_m(G)$ is {\em delay co-observable} w.r.t. $G$ and $\{\Sigma_{o,i}|i\in I\}$, if for all $s\in \bar K$, $u\in\Sigma_u^*$ with $su\in\bar K$ and for all $\sigma\in\Sigma_c$,
$su\sigma\in L(G)-\bar K\Rightarrow (\exists i\in I)P_{o,i}^{-1}(P_{o,i}(s))\Sigma_u^*\{\sigma\}\cap \bar K=\varnothing\wedge \sigma\in\Sigma_{c,i}.$}
\end{definition}

\begin{theorem}Given a language specification $K \subseteq L_{m}(G)$, for a plant $G$ with communication delays, there exists a nonblocking
decentralized supervisor $S_{dec}$ such that $L_{m}(S_{dec}/G) = K$ if and
only if
\begin{enumerate}
    \item $K$ is controllable w.r.t. $G$,
    \item $K$ is delay-coobservable w.r.t. $(\Sigma_{o,i}, \Sigma_{c,i})_{i \in \{1,2,\dots,n\}}$,
    \item $K$ is $L_{m}(G)$-closed.\hfill $\Box$
\end{enumerate}\end{theorem}

The computational complexity of verifying the delay-coobservability of a language $K$ is $O(|Q^K|^6|X|^2)$, where $Q^{K}$ is the state space of a deterministic automaton that recognizes $K$, and $X$ is the state set of $G$. 

Although the existence problem of the nonblocking supervisor has been solved in this paper, there are also some restrictions: 1) The assumption that all locally controllable events are locally observable is slightly restrictive; 2) The assumption that the number of possible subsequent occurrence of uncontrollable events is limited within a finite bound is a bit restrictive, as it excludes the possibility of a loop that contains an uncontrollable event in the plant; 3) This work only deals with the verification problem and cannot be used when the delay-observability property fails. Thus, it is of great importance to consider the synthesis problem when this property fails.

In \cite{tripakis2004decentralized} \cite{hiraishi2009solvability} \cite{sadid2015robustness} the authors take supervisor communication delays into consideration under the decentralized control architecture with communication among local supervisors. As a key issue in the supervisory control of networked DES, the property of decidability is investigated in \cite{tripakis2004decentralized} and \cite{hiraishi2009solvability}. In \cite{tripakis2004decentralized}, the problem of decentralized control with communication is studied, where delays are either bounded by a given constant,or unbounded. Communication channels are assumed to be FIFO and lossless. It is shown to be undecidable to check the existence of two controllers such that a set of responsiveness properties is satisfied, in both the case of unbounded-delay communication and the case of no communication. The decidability of joint observability with bounded-delay communication is also shown. By enforcing bisimilarity between the controlled system and the specification, the decentralized control problem is shown to be undecidable in \cite{hiraishi2009solvability}. This work also presents two sufficient conditions to make the decentralized control problem decidable for finite state controllers. The first condition is when communication involves $k$-bounded-delay, and the other is when any cycle in the state transition diagram of the system contains an event observable by all controllers. 

\subsubsection{$OC$ delays in modular/distributed control}
In \cite{zhang2016distributed} and \cite{zhang2016delay}, they start from the DES distributed control scheme called ``supervisor localization'', which describes a systematic top-down approach to design distributed controllers which collectively achieve global optimal and nonblocking supervision. Assuming that inter-agent communication of selected ``communication events'' may be subject to unknown time delays (no loss), a property of ‘delay-robustness’ is proposed and shown to be polynomial time verifiable, and that such tests serve to distinguish between events that are delay-critical and those that are not. In addition, timed DES is adopted as the system model in \cite{zhang2016delay} so that delays can be explicitly measured by the number of elapsed ticks. Then the property of timed delay-robustness with respect to the timed channel is defined, which extends the untimed counterpart. 

By modeling the modular systems as communicating finite state machines with reliable unbounded FIFO queues between subsystem, \cite{kalyon2011synthesis} and \cite{kalyon2013symbolic} adopt the technique of abstract interpretation for over-approximating reachability and co-reachabiility to ensure finite termination in distributed controller synthesis. As an extension of \cite{kalyon2011synthesis}, \cite{kalyon2013symbolic} provides the full process allowing to derive controllers from a state-based specification and a plant by means of state-based estimates and abstract interpretation techniques, whereas \cite{kalyon2011synthesis} only presents the control point of view with an overview of the state-based estimates computation point of view.

Based on distributed Petri net, \cite{darondeau2012distributed} proposes to synthesize distributed controller starting from a monolithic supervisor, where the communication channel is non-FIFO and lossless. By automatically encoding the information to be exchanged, which is a key contribution of this work, the distributed Petri net synthesis technique is illustrated on the 3 Dining Philosopher problem, producing three distributed solutions to it that could not have been discovered in the absence of an algorithmic strategy and a software tool.

\subsubsection{$OC$ delays in centralized control}
In \cite{alves2017supervisory}, the authors  consider non-FIFO observation channel with  delays and losses, while the control channel is assumed to be lossless and has no delay (thus effectively FIFO). The model of the timed networked discrete event systems (TNDES) is proposed, where an ordinary finite automaton model $G=(X, \Sigma, \delta,  x_0)$ of the plant is augmented with a timing structure $t: X \times \Sigma \longrightarrow \mathbb{R}^+$ to specify the minimal transition activation time. Then, an untimed one that models TNDES is recursively constructed. New properties of networked controllability and networked observability are defined and used to characterize the existence of a networked supervisor. All the relevant languages considered in \cite{alves2017supervisory} are prefix-closed, and thus the property of non-blockingness is not studied. Compared with the timed discrete event systems (TDES) model, the number of states used for representation of a system may be significantly reduced by using TNDES. However, a model transformation to an untimed model still needs to be carried out. This may reduce the benefit of using TNDES. The number of transitions of the untimed model is exponential in the size of the alphabet.

In \cite{alves2019robust}, the authors consider the problem of design of robust supervisors that are able to cope with intermittent loss of observations and also make the controlled system achieve the specification language under nominal operation. Necessary and sufficient conditions, i.e., robust controllability and $K$-observability, for the existence of a robust supervisor that is able to cope with intermittent loss of observations is presented. The property of relative observability is also extended to robust relative observability. Furthermore, the verification conditions of robust controllability and $K$-observability extends classical Ramadge-Wonham controllability and observability properties. However, there is no consideration of delays in the communication channels, which may make the work unrealistic. The problem of synthesis of a robust supervisor when the characterizing conditions fail is not studied.

\subsubsection{Verification and detectability related to $OC$ delays}
By modelling the system as a timed DES where an explicit tick event is used to measure the passage of one unit of time, \cite{sadid2015robustness} verifies the robustness of all synchronous communication protocols under conditions of fixed or finitely-bounded delay (no loss), not just optimal communication protocols. However, this work only addresses the problem of verification of robustness against delay; if the communication protocol is not robust against delays, then the theory of this work cannot be used.

In \cite{zhou2019supervisory}, the authors consider the problem of observation nondeterminism. A method is proposed to check $O$-observability by constructing an augmented automaton. In the augmented automaton, each state is a doubleton of which the first element is the current state of the original system (which is used to track all the strings generated by the original system) and the second element is a set of state estimates of all the possible observations when a strings occurs. A subset is defined that includes all the bad states in the augmented automaton. It is shown that $O$-observability holds if and only if there are bad states in the augmented automaton. A state-estimate-based supervisor is synthesized to control the given system to be safe when the nondeterministic control problem is solvable.

In \cite{sasi2018detectability}, the authors study the detectability for networked discrete event systems impacted by network delays and losses, which is concerned with the ability to determine the current and subsequent states. This work considers both network detectability and network D-detectability. Network detectability allows the determination of the state of a networked discrete event system, while networked D-detectability allows one to distinguish between some pairs of states of the systems. The characterization and verification of these two detectability properties are also provided in \cite{sasi2018detectability}.

The problem of state estimation under communication delays, for non-FIFO channel, has been considered in \cite{lin2019state}. It considers multiple channels, each of which is a FIFO channel with a different delay. Thus, the resulting channel is non-FIFO. The first, conservative, method for computing state estimate is directly extended from an existing approach, which computes an over-approximation in the sense that this state estimate may contain states that the system cannot be in. The second, exact, method distinguishes between the occurrence of an event and the reception of an event, as in the general discussions in the beginning of this subsection. Each communication channel is modeled with these two types of events. State estimates can be computed based on the synchronous product of the plant model and the channel models. Both online computation and offline computation methods are proposed. This work extends the state estimate method when no communication delay is involved.

In \cite{alves2019state}, both communication losses and delays in the observation channel are considered in this work. The observation channel is assumed to be non-FIFO, since there are multiple FIFO observation channels. The plant is in general a non-deterministic finite automaton, and delay is measured by the number of occurrences of events. By a transformation to untimed nondeterministic automaton, networked $D$-detectability definition is proposed, which is equivalent to the $D$-detectability in the untimed nondeterministic model. The $D$-detectability properties studied include strong $D$-detectability, weak $D$-detectability, strong periodic $D$-detectability, weak periodic $D$-detectability. 

\subsection{State-of-art networked control with $OC$ and $CC$ delays}
In this subsection we organize existing publications based on specific models of channel delays and relevant control architecture. 
 
\subsubsection{An input-output control architecture with non-FIFO channels}
\begin{figure}[htb]
    \begin{center}
      \includegraphics[height=0.4\textwidth, width=0.4\textwidth]{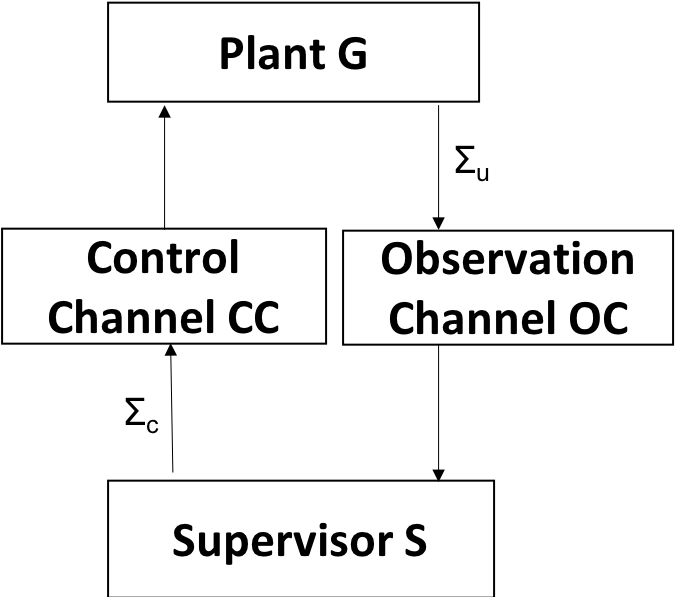}
    \end{center}
    \caption{An input/output interpretation of supervisory control of networked discrete-event systems}
    \label{fig:reinterACC}
\end{figure}

In \cite{balemi1992supervision}, the authors present an input-output interpretation of supervision of discrete-event systems.  In their networked setup, the schematic diagram is shown in Fig.~\ref{fig:reinterACC}, 
where the supervisor sends the commands in  $\Sigma_c$ through the control channel and the plant sends the responses in $\Sigma_u$ to the supervisor.
More specifically, the plant produces responses in reaction to commands from the supervisor and, symmetrically, the supervisor accepts the responses of the plant and produces commands for the plant. In this input-output perspective, each command from the supervisor is a controllable event, while each response from the plant is an uncontrollable event. \cite{balemi1992supervision} makes the  assumption that the observation channel can hold multiple responses in $\Sigma_u$ and the control channel can only hold one command in  $\Sigma_c$. The plant is given by $G=(\Sigma, L_G, M_G)$, where $L_G$ denotes the closed-behavior of $G$ and $M_G \subseteq L_G$ denotes the marked behavior of $G$. Similarly, the supervisor is given by $S=(\Sigma, L_S, M_S)$. The composition of $G$ and $S$ is denoted by $G \lVert S=(\Sigma, L_G^c, M_G^c)$. 
The composition $G\lVert S$ of $G$ and $S$ is said to be well-posed if $G\lVert S=G \lVert_{(\Sigma_u, \Sigma_c)} S$, where $\lVert_{(\Sigma_u, \Sigma_c)}$ denotes the prioritized synchronous composition operator w.r.t. $(\Sigma_u, \Sigma_c)$. In their networked supervisor  synthesis problem formulation, ~\cite{balemi1992supervision} requires a) $\varnothing \subset P_{\Sigma_u}(M_G^c) \subseteq L_{spec}'$, where $L_{spec}'\subseteq \Sigma_u^*$ denotes the specification and $P_{\Sigma_u}: \Sigma^* \longrightarrow \Sigma_u^*$ denotes the natural projection, b) $G$ and $S$ is well-posed, c) $S$ is non-blocking in the closed-loop and a marking in $S$ eventually corresponds to a marking in $G$, and d) the behavior of $S$ is unaffected by permutation of order of commands and responses in $G$. To solve the synthesis problem, the notion of a delay-insensitive language is proposed in \cite{balemi1992supervision}.
The main theorem that characterizes  the existence of a delay-insensitive supervisor $S$ is given in the following. \begin{theorem}
\label{theorem: delay}
There exists a delay-insensitive supervisor $S$ for $G$ such that $M_G^c=K$ iff $K$ is controllable, delay-insensitive and $M_G$-closed.
\end{theorem}
It turns out that, although the class $D(L)$ of delay-insensitive sublanguages of a language $L$ is not closed under union, the class $CD(L)$ of delay-insensitive and controllable sublanguages of $L$ is closed under union, and thus the supremal element $sup CD(L)$ exists in $CD(L)$.  Thus, the following result holds, which allows one to solve the synthesis problem even if the above characterizing conditions fail.
\begin{theorem}
The supervisory control problem with delays is  solvable if and only if $\varnothing \subset P_{\Sigma_u}(sup CD(M_G \cap P_{\Sigma_u}^{-1}(L_{spec}')))$. If a solution exists, then the supervisor $S=(\Sigma, \overline{K_D}, K_D)$ with $K_D=sup CD(M_G \cap P_{\Sigma_u}^{-1}(L_{spec}'))$ is the maximally permissive solution.
\end{theorem}
Another important result is the following, which allows a networked supervisor to be synthesized as in the case of delay-free communications. 
\begin{theorem}
\label{theorem:C}
If $M_G$ and $L_{spec}'$ are delay-insensitive, then $K=sup C(M_G \cap P_{\Sigma_u}^{-1}(L_{spec}'))$ is also delay-insensitive.
\end{theorem}
In the extended work \cite{balemi1994input}, the effect of communication delays on the connection of a plant and a supervisor is defined via a delay operator. Formally, the delay of a language $L \subseteq \Sigma^*$ with respect to $\Sigma' \subseteq \Sigma$ is denoted by $delay[\Sigma'](L)$, which is defined to be the smallest superlanguage of $L$ such that for any $s, t \in \Sigma^*, \sigma' \in \Sigma', \sigma \in \Sigma-\Sigma'$
\begin{enumerate}
    \item $L \subseteq delay[\Sigma'](L)$
    \item $s\sigma'\sigma t \in delay[\Sigma'](L) \implies s\sigma \sigma' t\in delay[\Sigma'](L)$
\end{enumerate}
Intuitively, $delay[\Sigma'](L)$ is the closure of $L$ under the permutation of  the substrings $\sigma'\sigma\in \Sigma'(\Sigma-\Sigma')$ to $\sigma\sigma'\in (\Sigma-\Sigma')\Sigma'$. A notion of well-posedness is defined, which informally requires that 
any command coming from the supervisor to the plant must be accepted by the plant; the continuation of the current string in the plant with a command from the supervisor must form a new string contained again in the closed-behavior $L_G$ of the plant. The class $\omega[L_G](L)$ of sublanguages of $L$ for supervisors enforcing well-posedness of the connection with a plant having language $L_G$ is shown to be closed under union; the supremal element is denoted by  $sup \omega[L_G](L)$.  The unmarked supervisor synthesis problem with delays is formulated in \cite{balemi1994input} as follows. Given a plant $G=(\Sigma, L_G, L_G)$ and a prefix-closed specification language $L_{spec}' \subseteq \Sigma_u^*$, find a supervisor $S=(\Sigma, L_S, L_S)$ such that a) $L_S \subseteq L_G$, b) $\varnothing \subset P_{\Sigma_u}(L_G^c) \subseteq L_{spec}'$, and c) the connection of $S$ and $G$ is well-posed. It is shown that the above unmarked supervisor synthesis problem with delays, where all relevant languages are prefix-closed, has a solution if and if for the language  $L_S=sup\{K \mid K \in \omega[L_G](L_G)\wedge \overline{L_G \cap delay[\Sigma_c](K)} \subseteq P_{\Sigma_u}^{-1}(L_{spec}')\}$, it holds that $\varnothing\subset P_{\Sigma_u}(L_S)$. If a solution exists, then the supervisor with language $L_S$ is a solution. To ensure computability, the language $L_G$ is required to be self-well-imposed, that is, $L_G \supseteq L_G\Sigma_c^* \cap \overline{delay[\Sigma_c](L_G\Sigma_u^*})$ and $L_G \supseteq L_G\Sigma_u^* \cap \overline{delay[\Sigma_u](L_G\Sigma_c^*})$. If $L_G$ is self-well-imposed, then the above $L_S$ can be replaced with $supC(L_G\cap P_{\Sigma_u}^{-1}(L_{spec}'))$ (cf. Theorem~\ref{theorem:C}). The restriction to plants with so called memoryless languages allows one to compute in polynomial time a supervisor solving the networked supervisor synthesis problem with communication delays. A prefix-closed language $L\subseteq \Sigma^*$ is said to be memoryless if for any $s, s' \in L$ such that $s \in delay[\Sigma_c](\{s'\})$, then for any $t\in \Sigma^*$, $st\in L \iff s't \in L$. Moreover, all the characterizing results have also been extended to dealing with non-prefix-closed cases in
\cite{balemi1994input} to address the non-blockingness property. It is argued that most systems can be properly modeled to satisfy the memoryless property and another technical condition to allow polynomial time synthesis of networked supervisors. How this can be achieved in practice is not discussed in detail.

\subsubsection{An implicit model of $OC$ and $CC$ delays in centralized control}
\begin{figure}
    \centering
    \includegraphics[scale=0.5]{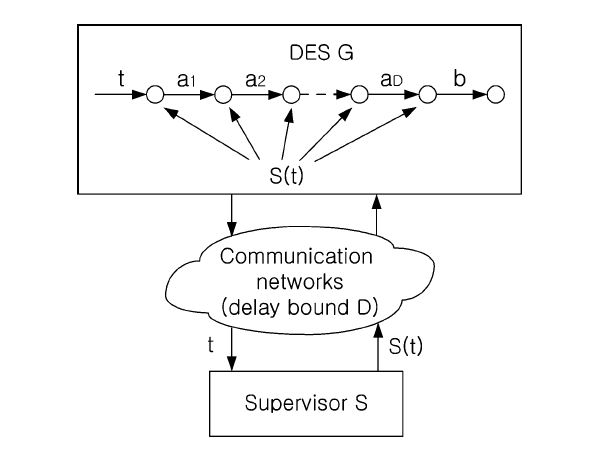}
    \caption{Supervisory control of a DES subject to a delay bound D}
    \label{fig:Supervisory control of a DES subject to a delay bound D}
\end{figure}

In \cite{park2006delay}, the authors  investigate the existence conditions of a delay-robust non-blocking supervisor that can achieve  a given language specification, for FIFO observation channel and control channel with bounded communication delays. The schematic diagram is shown in Fig.~\ref{fig:Supervisory control of a DES subject to a delay bound D}, where the delay bound is assumed to be $D$. Following the general discussion, it is assumed that a control pattern only contains controllable events, and every controllable event is disabled by default and is permitted to occur only if it is enabled by a supervisor. Thus, a supervisor is a map $S:L(G)\longrightarrow 2^{\Sigma_c}$. According to the schematic diagram,  uncontrollable events may subsequently occur within a delay bound $D$ (from the moment the supervisor sends a control pattern), and thus the supervisory control action $S(t)$ for a string $t$ can be actually applied to the system either after the string $t$ or further after any subsequent occurrence of uncontrollable events $a_1...a_i$ where $i\in \{1,...,D\}$. The closed-loop behavior $L(S/G)$ is defined as follows: 1) $\epsilon \in L(S/G)$, and 2) for any $s\in L(S/G)$ and $\sigma \in \Sigma$ with $s\sigma \in L(G)$, $s\sigma\in L(S/G) \iff i) \sigma\in \Sigma_{u}$, or, ii) $\exists t \in \overline{s}, |s|-|t|\leq D$ and $\sigma \in S(t)$  whereas $tu=s$ for some $u\in \Sigma_u^*$ and $S(tv)=\varnothing$ for any $v\in \overline{u}-\{\epsilon\}$. For a specification $K \subseteq L_m(G)$ for $G$ subject to a delay bound $D$, the problem is to find necessary and sufficient conditions for the existence of a nonblocking supervisor $S$ such that $L_m(S/G) = K$, where as usual $L_m(S/G)=L(S/G)\cap L_m(G)$ and $S$ is nonblocking iff $\overline{L_m(S/G)}=L(S/G)$. To solve the problem, it is assumed that every possible subsequent occurrence of uncontrollable events
is limited within the delay bound $D$, i.e., $|u|\leq D$ for any $u \in \Sigma_u^+$ and $s \in \Sigma^*$ satisfying $su \in L(G)$. Based on this assumption, the property of delay-nonconflictingness is formulated and the following characterization result is proved.
\begin{theorem}
Given a specification $K \subseteq L_m(G)$ for $G$ subject to a delay bound $D$, there exists a delay-robust nonblocking supervisor $S$ such that $L_m(S/G)=K$ iff 
\begin{enumerate}
    \item $K$ is controllable w.r.t. $G$ and $\Sigma_u$
    \item $K$ is delay-nonconflicting w.r.t. $G$ and $D$
    \item $K$ is $L_m(G)$-closed.
\end{enumerate}
\end{theorem}
\noindent 
Suppose $G$ is modeled by a finite-state automaton $G = (Q,\Sigma,q_0,\delta,Q_m)$ and $K$ is modeled by  $H = (\Sigma,X,\xi,x_0,X_m)$. Then, an algorithm is provided to perform the verification of the delay-nonconflictingness of $K$, which has a computational complexity of $O(|X||Q||\Sigma_{u}|^D)$. 

There are two limitations of this work. Firstly, there is no synthesis algorithm provided if the delay-nonconflictingness property fails. Secondly, it is assumed that every plant event is observable, which may not be realistic in practice. In \cite{park2007supervisory}, the authors extend the characterization results of their previous work \cite{park2006delay} to the case of partial observation, by proposing the notion of delay observability to replace the notion of delay-nonconflictingness defined for the full observation case. The delay observability property also  assures no confliction  in making a decision for legal controllable events under partial observation and communication delays. The setup assumes controllable events are observable, and it only addresses the verification problem, and thus the theory cannot be used if the verification conditions fail. 

In \cite{lin2014control}, the author considers the problem of networked supervisor synthesis, dealing with both communication delays and losses. In particular, delays and losses occur in both the observation channel and the control channel, which are assumed to be FIFO. For the case when the control channel has no delays or losses, two observation maps are used for dealing with observation delays and losses, respectively.
Let $G=(Q, \Sigma, \delta, q_0)$. Now, let $\delta_o=\{(q, \sigma, q')\mid \delta(q, \sigma)=q' \wedge \sigma \in \Sigma_o\}$ denote the set of observable transitions and let $\delta_{uo}=\{(q, \sigma, q')\mid \delta(q, \sigma)=q' \wedge \sigma \in \Sigma_{uo}\}$ denote the set of unobservable transitions. Let $\delta_L \subseteq \delta_o$ denote the subset of observable transitions that may get lost in the communication. Let $\Theta_L: L(G) \longrightarrow 2^{\Sigma_o^*}$ denote the observation mapping under losses, which is defined as follows: let $s=\sigma_1\ldots\sigma_i\ldots s_k \in L(G)$, then $\Theta_L(s)$ is obtained by replacing each $\sigma_i$ in $s$ with a) $\{\epsilon\}$, if the corresponding transition $(q_i, \sigma_i, \delta(q_i, \sigma_i))\in \delta_{uo}$, b) $\{\sigma_i\}$, if $(q_i, \sigma_i, \delta(q_i, \sigma_i))\in \delta_o-\delta_L$ and c) $\{\epsilon, \sigma_i\}$, if $(q_i, \sigma_i, \delta(q_i, \sigma_i))\in \delta_L$. That is, network losses lead to nondeterminism in observation and thus a string $s$ may lead to a set $\Theta_L(s)$ of observed strings.  $\Theta_L(G)$ is recognized by $G_L=(Q, \Sigma_o, \delta_{loss}, q_0)$, where $\delta_{loss}=\delta_o\cup \{(q, \epsilon, q') \mid (q, \sigma, q')\in \Sigma_{uo}\cup \delta_L\}$. That is, $\delta_{loss}$ is obtained from $\delta$ by adding  transitions $(q, \epsilon, q')$, whenever $(q, \sigma, q')\in \delta_L$ for some $\sigma \in \Sigma_o$. To model $N$-bounded delays, $\Theta_{D}^N:\Sigma^* \longrightarrow 2^{\Sigma^*}$ is used for delayed observation. Formally, for any $s\in L(G)$, $\Theta_{D}^N(s):=\{s_{-i} \mid i \in [0, N]\}$, where $s_{-i}$ is the prefix of $s$ with the last $i$ events removed. Intuitively, $s_{-i}$ may be observed after $s$ is executed, since the last $i$ events in $s$ may not be observed yet. With both observation delays and losses, the map  $\Theta_{DL}=\Theta_L \circ \Theta_{D}^N$  is defined.  The network observability is defined as follows, which is reduced to observability  if $\Theta_{DL}(s)=P_o(s)$ in the setup with no delays and no losses.
\begin{definition}
Given a prefix-closed language $K \subseteq L(G)$ and the observation mapping
under communication delays and losses described by $\Theta_{DL}$ with a delay upper bound $N$, $K$ is network observable with respect to $L(G)$ and $\Theta_{DL}$ if $\forall s\sigma \in L(G)$, $s\sigma \in K \Longrightarrow \exists t \in \Theta_{DL}(s), \forall s'\in \Theta_{DL}^{-1}(t), s' \in K \wedge s'\sigma \in L(G) \Longrightarrow s'\sigma\in K$.
\end{definition}
The author considers a state-estimate based networked supervisor $\gamma: 2^Q \longrightarrow 2^{\Sigma}$ that determines for each state estimate the set of events to be disabled. The state estimate of the supervisor after observing $t \in \Sigma_o^*$ is $E(t)=\{q \in Q \mid \exists s\in L(G), t \in \Theta_{DL}(s) \wedge \delta(q_0, s)=q\}$. 
Then, the closed behavior $L(G, \gamma)$ of the supervised system  is defined recursively as follows: a) $\epsilon \in L(G, \gamma)$, and b)$\forall s\in L(G, \gamma)$, $\forall \sigma \in \Sigma$,  $s\sigma \in L(G, \gamma)$ iff $s\sigma \in L(G) \wedge (\sigma \in \Sigma_{u} \vee \exists t \in \Theta_{DL}(s)\sigma \notin \gamma(E(t)))$. The following characterization result is obtained.
\begin{theorem}
Assume a networked discrete event system $G$ with communication
delays and losses in observation, described by $\Theta_{DL}$ with an upper bound $N$. Assume
that there are no communication delays or losses in control. For a nonempty prefix-closed
regular language $K \subseteq L(G)$, there exists a state estimate-based networked supervisor $\gamma: 2^Q \longrightarrow 2^{\Sigma}$ such that $L(G, \gamma) = K$ if and only
if (1) $K$ is controllable with respect to $L(G)$ and $\Sigma_u$ and (2) $K$ is network observable
with respect to $L(G)$ and $\Theta_{DL}$.
\end{theorem}
The author then considers the control channel with communication delays upper  bounded by $M$ and $M$-bounded consecutive losses of control patterns, when the observation channel has no losses or delays and under full observation. Let $K/s:=\{s' \in \Sigma^* \mid ss'\in K\}$. Then, network controllability is defined in the following, which is reduced to controllability when $M=0$.
\begin{definition}
Given a prefix-closed language $K \subseteq L(G)$ and an upper bound $M$ on
control delays and losses,  $K$ is said to be network controllable with respect to $L(G)$
and $\Sigma_u$ if $\forall s \in K, \forall \sigma \in \Sigma, s\sigma \in L(G) \wedge (\sigma \in \Sigma_u \vee \sigma \in K/s_{-1}\vee \sigma \in K/s_{-2}\vee \ldots \sigma \in K/s_{-M})\Longrightarrow s\sigma \in K$.
\end{definition}
Unlike  controllability, even if all events are controllable,
$K$ may not be network controllable, due to communication delays and
losses in control. It is assumed that  before a control action arrives, the system will use the previously
received control action. Then, the closed behavior $L(G, \gamma)$ of the supervised system  is defined recursively as follows: a) $\epsilon \in L(G, \gamma)$, and b)$\forall s\in L(G, \gamma)$, $\forall \sigma \in \Sigma$,  $s\sigma \in L(G, \gamma)$ iff $s\sigma \in L(G) \wedge (\sigma \in \Sigma_{u} \vee \sigma \notin \gamma(\delta(q_0, s)) \cap \gamma(\delta(q_0, s_{-1})) \cap \ldots \cap \gamma(\delta(q_0, s_{-M})))$. The following characterization result is obtained.
\begin{theorem}
Assume a networked discrete event system $G$ with communication
delays and losses in control, bounded by $M$. Assume full observation and that there are no communication delays or losses in observation. For a nonempty prefix-closed regular
language $K \subseteq L(G)$, there exists a state-based
networked supervisor $\gamma: Q \longrightarrow 2^{\Sigma}$ such that $L(G, \gamma) = K$ if and only if $K$ is network
controllable.
\end{theorem}
In the general case, when the control channel has both delays and losses bounded by $M$ and the observation channel has both delays and losses, with upper bound $N$, a principle of separation holds. Thus, the following central result holds.
\begin{theorem}
Assume a networked discrete event system $G$ with communication
delays and losses in observation, described by $\Theta_{DL}$ (upper bound $N$), and with communication delays and losses in control bounded by $M$. For a nonempty prefix-closed regular
language $K \subseteq L(G)$, there exists a state-estimate based networked supervisor $\gamma: 2^Q \longrightarrow 2^{\Sigma}$ such that $L(G, \gamma) = K$ if and only if (1) $K$
is network controllable with respect to $L(G)$ and $\Sigma_u$ and (2) $K$ is network observable
with respect to $L(G)$ and $\Theta_{DL}$.
\end{theorem}
 While the non-networked supervisory control framework has been naturally  extended to the networked setup in \cite{lin2014control}. This work only studies the verification conditions for characterizing the existence of a networked supervisor that achieves a given specification language. If network observability or network controllability fails, then a supervisor cannot be synthesized according to the results of this work.
In \cite{shu2014supervisor}, the authors consider a slightly different problem setup where the behavior of the supervised system needs to be both adequate and safe. 
The observation channel is assumed to have a delay bound $N^o$ and the control channel has a delay bound $N^c$.  Any  control policy $\pi$ has to satisfy both the observation feasibility and control feasibility. A supervisor may disable different events for the same sequence of event occurrences in the plant, due to the random nature of the delays. Thus, to describe the behavior of
the controlled system,  two languages are defined in \cite{shu2014supervisor}. The first language, which is smaller, contains strings that can
be generated under all observation and control delays and is denoted by
$L_r(\pi/\hat{G})$, where $\hat{G}$ denotes the networked discrete-event system consisting of the plant $G$ and the channel delays. The second language, which is larger, contains strings that
can be generated under some possible observation and control delays
and is denoted by $L_a(\pi/\hat{G})$. The synthesis problem, i.e., the Supervisor Synthesis Problem for Networked Discrete Event Systems, is then formulated as follows: given a networked discrete event system $\hat{G}$, a
minimal required language $K_r$, and a maximal admissible language
$K_a$, synthesize a supervisor with control policy $\pi$ such that: 1) $\pi$ is
control feasible; 2) $\pi$ is observation feasible; 3) $L_r(\pi/\hat{G}) \supseteq K_r$; and
4) $L_a(\pi/\hat{G}) \subseteq K_a$. It is shown in \cite{shu2014supervisor} that the class of control feasible and observation feasible control policies is closed under conjunction. Furthermore, $L_r(\pi_1\wedge \pi_2/\hat{G})=L_r(\pi_1/\hat{G})\cap L_r(\pi_2/\hat{G})$. Thus, there exists the minimal policy $\pi_{min}$ such that $L_r(\pi_{min}/\hat{G}) \supseteq K_r$. The following main characterization result is then obtained.
\begin{theorem}
The Supervisor Synthesis Problem for Networked Discrete
Event Systems is solvable if and only if $L_a(\pi_{min}/\hat{G})\subseteq  K_a$. Furthermore, if it is solvable, then $\pi_{min}$ is a solution.
\end{theorem}
\cite{shu2014supervisor} also constructs $\pi_{min}$ and proposes an implementation of $\pi_{min}$ based on a  state-estimated based control policy and a new observer. The off-line implementation of the minimal supervisor  is of exponential complexity, while an on-line implementation can reduce the computational complexity to be
polynomial in each step. An  algorithm to check the condition $L_a(\pi_{min}/\hat{G})\subseteq  K_a$ for the existence of a networked supervisor is proposed by constructing an augmented automaton $G_{aug}$. A maximally-permissive control policy $\pi_{max}$ can be obtained from $\pi_{min}$ by an iterative construction.
Compared with \cite{lin2014control}, the results developed in \cite{shu2014supervisor} do not need to assume that a uniform delay is applied to all the events delayed in a string generated by the plant. However, due to the random delays, the language generated by the controlled system is nondeterministic, which makes it more difficult to analyze  properties such as nonblockingness/deadlock-freeness.  Even if  a  string has  a  continuation, the  controlled  system  can  still  be blocked/deadlocked after that string, as the continuation may be disabled in some (but not all) trajectories the controlled system may take due to nondeterminism.  In~\cite{shu2016deterministic}, the authors explicitly address this nondeterminism. In order to capture the nondeterminism caused by the communication delays in the observation channel,  delay observability is defined which says that if two event sequences have different control requirements, then all  the  possible observations  of  them  must  be  totally  different.  If  there  are  no communication delays, then delay observability is reduced to observability. In order to capture the nondeterminism caused by the communication delays in the control channel, delay controllability is defined. Delay controllability says that if one needs to disable an event, then that event must be controllable and all the possible controls must disable it. If there are no communication delays, then delay controllability is reduced to controllability.
Deterministic Networked Control Problem for Discrete Event Systems is formulated as follows: given a plant $G$ subject to observation delays and control delays, and a specification language $K$, find an observation feasible and control feasible control policy  $\pi$ such that the controlled system $\pi/G$ satisfies $L_r(\pi/G)=L_a(\pi/G)=K$. It turns out that the Deterministic Networked Control Problem for Discrete Event Systems is solvable iff $K$ is delay controllable and the augmented language $K^{aug}$ is delay observable.
Algorithms are also proposed in \cite{shu2016deterministic} to verify these two properties.  If the language to be synthesized is not delay observable and/or delay controllable, its infimal delay controllable and delay observable  superlanguage  and  maximal  delay  controllable  and delay  observable  sublanguages are also constructed in \cite{shu2016deterministic}. In \cite{wang2016robust}, the authors study the robust control of networked discrete-event systems, where a supervisor is used to control several possible plants under communication delays and losses. The solution methodology is by translating the robust control problem into the conventional networked control problem by constructing an augmented automaton for all possible plants and an augmented specification automaton for the corresponding specification automata. This work considers robust networked synthesis problem with both single objective and multiple objectives. The single objective case corresponds to when all the specifications are the same. A necessary and sufficient condition for the existence of a robust networked supervisor is derived. The multiple objectives case corresponds to when the specifications are different. Only a sufficient condition for the existence of a networked supervisor is obtained. In \cite{shu2016predictive}, the authors extend the work of \cite{lin2014control} to consider predictive networked supervisor, which predicts the impacts of communication delays and losses in the control channel in determining the control actions. The existence condition of a predictive networked supervisor is derived, based on controllability and network observability. It is shown that predictive networked supervisors are better than non-predictive counterparts and predictive networked supervisor is optimal in the sense that it is always a solution if the networked supervisory control problem is solvable.

In \cite{zhao2015supervisory}, the authors present an application of control of networked timed discrete event systems to power distribution networks.  Under the assumption that delays and losses are bounded, a necessary and sufficient condition based on network T-controllability and network T-observability are used to characterize the existence of networked supervisor. The results are applied to 33-node (bus) test system, where the objective is to ensure that the total sub-station transformer power stays within prespecified safety limits. In \cite{hou2019relative}, the authors introduce and reduce relative network observability, under  communication delays and losses in the FIFO observation channel and control channel, to network observability, which allows existing solutions for network observability verification to be directly applied; the application to the calculation of the supremal controllable and relatively network observable sublanguage is also shown.  

\subsubsection{An implicit model of $OC$ and $CC$ delays in decentralized control}
Following the channel delay model proposed in \cite{lin2014control}, the authors in \cite{shu2014decentralized} discuss decentralized control and investigate how to use the local supervisors to control the system in order to satisfy given specifications under the influence of both $OC$ and $CC$ delays. The specifications are described by two languages: a minimal language specifying the minimal required performance to achieve, and a maximal admissible language specifying the maximal set of legal behaviours. This work is an extension of a centralized framework described in \cite{shu2014supervisor} to decentralized networked control setting, assuming each local supervisor has its own communication channel with the plant and different communication channels may have different communication delays.
It is assumed that in the $OC$, communication delays do not change the order of the events, i.e., the observation channel is FIFO. In the control channel, the initial control policy is not delayed. 

Due to observation delays, local supervisors may have different observations for the same string $s\in L(G)$. By adopting the similar mapping function to capture the relationship between the string observed by local supervisors and the string generated by the plant, the set of possible observations for local supervisor $S_i$ ($i\in I$) is denoted by $\Theta_i(s) = \{P_{o,i}(t)|(\exists m \in \{0,\cdots, N_{o,i}\}) t = s_{-m}\}$, where $N_{o,i}$ is the upper bound of delays in the $OC$ and $s_{-m}$ is the prefix of $s$ obtained by removing the last $m$ events. $\theta_i(s)\in \Theta_i(s)$ is used to denote a particular (delayed) observation. The control policy $\pi_i$ implemented by local supervisor $S_i$ should be calculated based on the current observation, that is, $\pi_i: \Sigma^* \times \Sigma_{o,i}^* \rightarrow \Gamma$. This paper adopts the conjunctive fusion rule to combine control actions of local supervisors. The decentralized control map is $\hat{\pi}:\Sigma \rightarrow\Gamma$, where 
\[(\forall s\in L(G))\hat \pi(s):=\bigcap_{i\in I}\pi_i(s,\theta_i(s)).\]
The closed-loop system is defined as $\hat{\pi}/G$ in a usual way. The decentralized control problem is stated as follows.

\begin{problem} 
\emph{(DCPNDES)}: Given a plant $G$, a minimal required prefix language $K_r$ recognized by $G_r = (Q_r, \Sigma, \delta_r, q_0, Q_r)$, and a maximal admissible prefix language $K_a$ recognized by $G_a = (Q_a, \Sigma, \delta_a, q_0, Q_a)$, we want to find a decentralized control policy $\hat{\pi}$ such that
\begin{enumerate}
    \item $\hat{\pi}$ is co-control feasible, that is, \[(\forall i\in I)(\forall s\in L(G))\, \Sigma_u\subseteq\pi_i(s,\theta_i(s));\]
    \item $\hat{\pi}$ is co-observation feasible, that is, \\
    $(\forall i\in I)(\forall s,s'\in L(G))\theta_i(s)=\theta_i(s')\Rightarrow \\ \pi_i(s,\theta_i(s))=\pi_i(s',\theta_i(s'));$
    \item $K_r\subseteq L_r(\hat{\pi}/G)$, where 
    \begin{itemize}
    \item $\epsilon\in L_r(\hat\pi/G)$,
    \item $s\sigma\in L_r(\hat\pi/G)\iff s\in L_r(\hat\pi/G)\wedge s\sigma\in L(G)\wedge (\forall i\in I)(\forall m_i\in \{0,\cdots, N_{c,i}\})(\forall \theta_i(s_{-m_i})\in\Theta_i(s_{-m_i}))\sigma\in \pi_i(s_{-m_i},\theta_i(s_{-m_i}))$.
    \end{itemize}
    \item $L_a(\hat\pi/G)\subseteq K_a$, where 
    \begin{itemize}
    \item $\epsilon\in L_a(\hat\pi/G)$,
    \item $s\sigma\in L_a(\hat\pi/G)\iff s\in L_a(\hat\pi/G)\wedge s\sigma\in L(G)\wedge (\forall i\in I)(\exists m_i\in \{0,\cdots, N_{c,i}\})(\forall \theta_i(s_{-m_i})\in\Theta_i(s_{-m_i}))\sigma\in \pi_i(s_{-m_i},\theta_i(s_{-m_i}))$. \hfill $\Box$
    \end{itemize}
\end{enumerate}
\end{problem}

To find a decentralized control policy $\hat{\pi}$ for DCPNDES, a minimal control policy is constructed for each local supervisor $S_i$ ($i\in I$). Assuming that, after a string $s\in L(G)$ occurs, the supervisor $S_i$ sees the string $\theta_i(s)$, the current state estimate of this supervisor is then given by
$E_{i}(\theta_{i}(s)) = \{q \in G_r|(\exists t \in L(G_r))\theta_i(s) \in \Theta_i(t) \wedge \delta_r(q_0, t) = q\}.$ 
For state $q$, the enabled event set under control delays is 
\[
\Gamma_{r}^{N_{c,i}}(q) = \bigcup\limits_{q^{'} \in R^{N_{c,i}}(q)}\Gamma_{r}(q^{'})
\]
where $\Gamma_{r}(q^{'})$ is the set of events defined at state $q^{'}$ in $G_{r}$, $N_{c,i}$ is the upper bound of delays in the control channel between $S_{i}$ and $G$, and 
$$
R^{N_{c,i}}(q) = \left\{
\begin{array}{ccl}
\{\delta_{r}(q,t)| t \in \Sigma^{*} \wedge |t| \leq N_{c,i}\}       &      & q \in Q_{r},\\
\varnothing     &      & otherwise.
\end{array} \right. 
$$
Then the minimal control policy $\pi_{i,min}(s, \theta_{i}(s))$ is given by
$$
\pi_{i,min}(s, \theta_{i}(s)) = \bigcup\limits_{q \in E_{i}(\theta_{i}(s))}\Gamma_{r}^{N_{c,i}}(q) \cup \Sigma_{u,i}
$$
where $\Sigma_{u,i}=\Sigma_u\cap\Sigma_i$ is the set of uncontrollable events for $S_i$. Based on the minimal control policy for each local supervisor, the decentralized conjunctive control policy $\hat{\pi}_{min}$ can be generated.
Finally, an augmented automaton similar to that in \cite{shu2014supervisor} is constructed to verify the existence of a solution.   
Although this work solves the supervisor synthesis problem for decentralized control, there are still some issues which could be investigated further: 
1) Since the observation channel is considered to be FIFO and no communication loss is considered in this work. It is of interest to consider non-FIFO channels and channels with communication loss; 
2) For the specific setup considered in this paper, it is significant to define the closed-loop behavior. A more difficult challenge is to remove the implicit requirement of a uniform delay, to make the framework more realistic; 
3) A proper definition of maximal permissiveness in the decentralized setup and its corresponding synthesis algorithm are also a point worthy of attention.

\subsubsection{An implicit model of $OC$ and $CC$ delays in modular control}
With the same channel delay model introduced in \cite{lin2014control}, the authors in \cite{komenda2016modular} consider the following modular control problem with $OC$ and $CC$ delays, which are finitely bounded respectively by $N_{o,i}$ and $N_{c,i}$.   
\begin{problem}\label{Prob4}
Given generators $G_1$ and $G_2$, whose alphabets are $\Sigma_1$ and $\Sigma_2$, respectively. Let $K\subseteq L(G_1||G_2)$ be a prefix-closed specification. Let $G_k=P_k(G_1)||P_k(G_2)$ be a properly designed coordinator and $K$ is conditionally decomposable with respect to $\Sigma_1$, $\Sigma_2$ and $\Sigma_k$ \cite{komenda2015coordination}, i.e., $K=P_{1+k}(K)||P_{2+k}(K)$, where $P_{i+k}:(\Sigma_1\cup\Sigma_2)^*\rightarrow (\Sigma_i\cup\Sigma_k)^*$ ($i=1,2$). Find two networked supervisors $S_1$ and $S_2$ with partial observation and $OC$ and $CC$ delays bounded by $N_{o,i}$ and $N_{c,i}$ such that
\begin{itemize}
\item $L(S_i/[G_i||G_k])\subseteq P_{i+k}(K)$;
\item $L(S_1/[G_1||G_k])||L(S_2/[G_2||G_k])=K$.\hfill $\Box$
\end{itemize}
\end{problem}
By decomposition $K$ into $P_{1+k}(K)$ and $P_{2+k}(K)$, the authors show the possibility to treat coordination and predictive control together to handle possible observation and command delays, which brings the advantage of lower computational complexity. By replacing $P_{o,i}$ with $\Theta_i$ associated with $N_{o,i}$ and $N_{c,i}$, the authors extend the concept of {\em conditional observability} from \cite{komenda2015coordination} to {\em conditional network observability}, and derive the following main result: Problem \ref{Prob4} is solvable if and only if
\begin{itemize}
\item $K$ is relaxed conditional controllable \cite{komenda2015distributed};
\item $K$ is conditionally network observable with respect to $N_{o,1}+N_{c,1}$ and $N_{o,2}+N_{c,2}$.
\end{itemize} 
The synthesis complexity is $O(2^{||P_{1+k}(K)||}+2^{||P_{1+k}(K)||})$, which in the worst case is double exponential-time, unless both $P_{1+k}$ and $P_{2+k}$ are natural observers \cite{wong2004computation}.


\subsubsection{An explicit automaton model of $OC$ and $CC$ delays}
\begin{figure}[htb]
    \begin{center}
 \includegraphics[height=0.4\textwidth, width=0.5\textwidth]{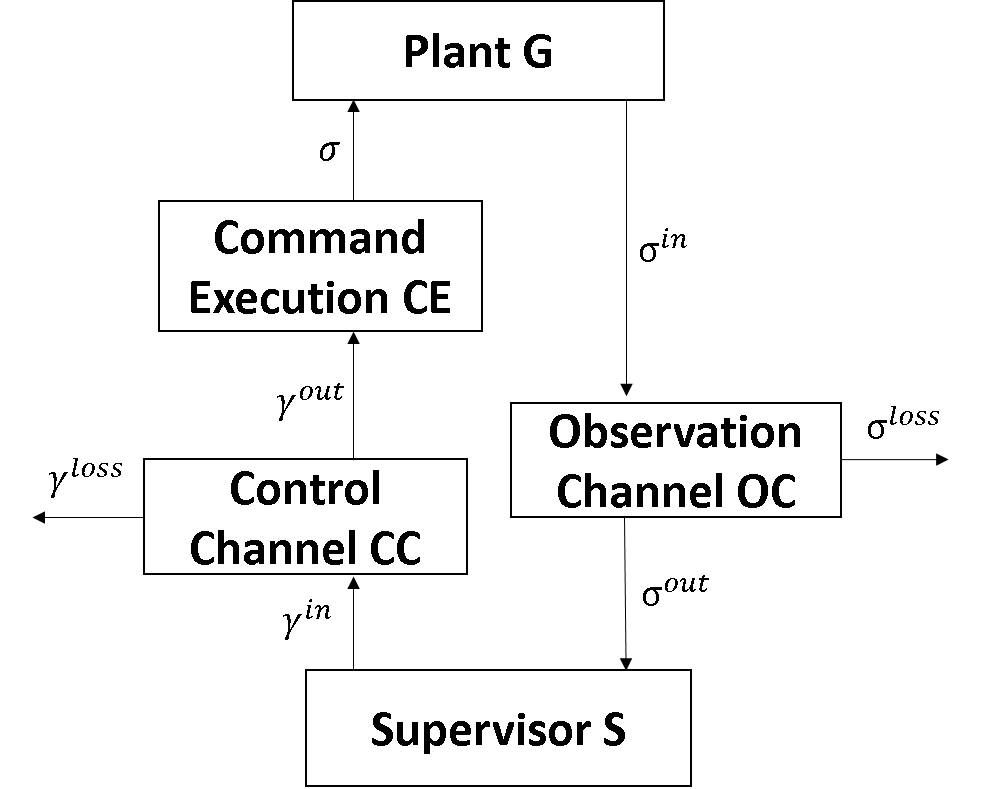}
    \end{center}
    \caption{A schematic diagram of a networked system with inputs and outputs}
    \label{fig:refined}
\end{figure}

In \cite{zhu2019supervisor} the authors present a new modeling framework, aiming to transform a networked control problem with $OC$ and $CC$ delays into a standard Ramadge-Wonham supervisory control problem. A schematic  diagram for a centralized setup is shown in Fig.~\ref{fig:refined}, which can be extended to a system of an arbitrary number of components and local supervisors.
Such a model transformation method makes it possible to apply existing supervisory control methods such as decentralized control, modular control and hierarchical control to networked control problems. Besides the four network components shown in Figure \ref{fig:diagram}, i.e., the plant $G$, the supervisor $S$, the observation channel $OC$ and the control channel $CC$, there is one extra component, the Command Execution Module (CE),  which is used to translate each control pattern $\gamma\in\Gamma$ into individual events so that the channel $CC$ model can be synchronized with the plant $G$. We will explain the details shortly.

The closed-loop system operates in the following way. Whenever the plant $G$ executes an observable event $\sigma \in \Sigma_o$, it sends a message $m_{\sigma}$, indicating the occurrence of $\sigma$ in the plant, over the observation channel; the event of sending the message $m_{\sigma}$ is denoted by $\sigma^{in} \in \Sigma_{o}^{in}$, where $\Sigma_{o}^{in}$ is a copy of $\Sigma_o$ with superscript ``$in$'', i.e., $\sigma^{in}\in\Sigma_o^{in}\iff \sigma\in\Sigma_o$. It is required that $\Sigma_o^{in}\cap\Sigma_o=\varnothing$. The event of receiving te message $m_{\sigma}$ by the supervisor is denoted by $\sigma^{out} \in \Sigma_o^{out}$, where $\Sigma_o^{out}$ is a copy of $\Sigma_o$ with superscript ``$out$'', and it is required that $\Sigma_o^{out}\cap\Sigma_o=\varnothing$. $\sigma^{out}$ can occur only if $\sigma^{in}$ has already occurred. When $\sigma^{in}$ occurs, but $\sigma^{out}$ never occurs, it is assumed that the message $m_{\sigma}$ is lost in transmission. In this case we use $\sigma^{loss}\in\Sigma_o^{loss}$ to denote message dropout. It is rquired that $\Sigma_o^{loss}\cap\Sigma_o=\varnothing$. In addition, to avoid the situation where each observable event may get lost, leading to no solution, we assume that only events in $\Sigma_{ol}\subseteq\Sigma_o$ may get lost. Each message $m_{\sigma}$ is characterised by the tuple $(\sigma^{in},\sigma^{out},\sigma^{loss})$.  The observation channel may have either a finite or an infinite capacity. After the supervisor $S$ receives an observation $\sigma^{out}$ from the observation channel, it sends a control command message $m_{\gamma}$ over the control channel, denoted by $\gamma^{in} \in \Gamma^{in}$, where $\Gamma^{in}$ is a copy of $\Gamma$ with superscript ``$in$'' such that $\Gamma^{in}\cap\Gamma=\varnothing$. Considering that $\Sigma_u$ is always allowed by the supervisor $S$, and execution of an uncontrollable event will be done autonomously by the plant $G$, thus, never be delayed in the $CC$, we assume that $\Gamma\subseteq 2^{\Sigma_c}$, that is, a control command only decides whether a controllable event should be disabled, denoted as the event not being included in the control pattern.  The event of receiving the message $m_{\gamma}$ by the plant $G$ is denoted by $\gamma^{out} \in \Gamma^{out}$, where $\Gamma^{out}$ is a copy of $\Gamma$ with superscript $``out"$ such that $\Gamma^{out}\cap\Gamma=\varnothing$. The control channel can have a finite or an infinite capacity and may also experience loss of messages. In  the case that the message $m_{\gamma}$ gets lost in the $CC$, the event $\gamma^{loss}\in\Gamma^{loss}$ will be used to denote the message dropout. We assume that $\Gamma^{loss}\cap\Gamma=\varnothing$. In addition, to avoid the situation where each control message may get lost, leading to no solution, we assume that only control patterns in $\Gamma_l\subseteq\Gamma$ may get lost. The $OC$ and $CC$ channel delays are respectively upper bounded by $num^o\in\mathbb{N}$ and $num^c\in\mathbb{N}$, which are interpreted as the number of event firings in the system, including both observable and unobservable events. 

Let $S_{obs}:=\{(m_{\sigma},i)|\sigma\in\Sigma_o\wedge 0\leq i\leq num^o\}$ be a set of all possible messages in the $OC$, where each message $m_{\sigma}$ is associated with a timer value $i$, which is upper bounded by $num^o$. When the timer value $i$ reaches 0, then the message $m_{\sigma}$ must either be popped out of the $OC$, or get lost. Nevertheless, the message $m_{\sigma}$ may be popped out before $i$ ticks down to 0, representing that the channel delay for $m_{\sigma}$ can be any value between 0 and $num^o$. Two assumptions are made below:
\begin{enumerate}
    \item The closed-loop system is asyncronous, i.e., no more than one event can fire at each time.
    \item The firing of each event in $\Sigma^{out}\cup\Sigma^{loss}$ does not trigger relevant timers to count down. 
\end{enumerate}

A non-FIFO $OC$ is modelled as a nondeterministic finite automaton 
$G_{OC}=(Q^{obs}:=2^{S_{obs}},\Sigma_{obs}:=\Sigma_o^{in}\cup\Sigma_o^{out}\cup\Sigma_o^{loss}\cup\Sigma_{uo},\delta^{obs},q_0:=\varnothing),$
where the transition map $\delta^{obs}:Q^{obs}\times\Sigma_{obs}\rightarrow 2^{Q^{obs}}$ is defined as follows: For each $q\in Q^{obs}$ and each $a \in \Sigma_{obs}$, let $\nu_q:=\max_{(m_{\sigma},i)\in q}i$, and $q'\in\delta^{obs}(q,a)$ if one of the following holds; otherwise, $\delta^{obs}(q,a)=\varnothing$:
\begin{enumerate}
\item $a=\sigma^{in}$, $\nu_q\geq 1$ and 
$q'=\{m_{\sigma'},i-1)\in S_{obs}|(m_{\sigma'},i)\in q\}\cup \{(m_{\sigma},num^o)\}$.
\item $a=\sigma^{out}$ and $q'\in \cup_{i:(m_{\sigma},i)\in q}\{q-\{(m_{\sigma},i)\}\}$.
\item $a=\sigma^{loss}$, $\sigma\in\Sigma_{ol}$, and $q'\in \cup_{i:(m_{\sigma},i)\in q}\{q-\{(m_{\sigma},i)\}\}$.
\item $a \in \Sigma_{uo}$, $\nu_q\geq 1$, and $q'=\{(m_{\sigma},i-1)\in S_{obs}|(m_{\sigma},i)\in q\}$.
\end{enumerate}
Intuitively, Rule
1 says that the firing of event $\sigma^{in}$ causes two consequences, i.e., the timer counts down by one unit for all messages in the channel and a new message $(m_{\sigma},num^o)$ is added into the channel. Rule 2 says that the firing of event $\sigma^{out}$ removes some message $(m_{\sigma},i)$ from the channel. Rule 3 says that some message $(m_{\sigma},i)$ with $\sigma\in\Sigma_{ol}$ may get lost. Rule 4 says that the firing of any unobservable event $\sigma \in \Sigma_{uo}$ will cause a timer countdown for all messages in the channel.  

A non-FIFO $CC$ can be treated in a similar way. Let $S_{com}:=\{(m_{\gamma},i)|\gamma\in\Gamma\wedge 0\leq i\leq num^c\}$ be a set of all possible messages in the $CC$, where each message $m_{\gamma}$ is associated with a timer value $i$, which is upper bounded by $num^c$. When the timer value $i$ reaches 0, then the message $m_{\gamma}$ must either be popped out of the $CC$, or get lost. Nevertheless, the message $m_{\gamma}$ may be popped out before $i$ ticks down to 0, representing that the channel delay for $m_{\gamma}$ can be any value between 0 and $num^c$. The $CC$ is modelled as a nondeterministic finite  automaton 
$G_{CC}=(Q^{com}:=2^{S_{com}},\Sigma_{com}:=\Gamma^{in}\cup\Gamma^{out}\cup\Gamma^{loss},\delta^{com} \linebreak ,q_0:=\varnothing),$
where the transition map $\delta^{com}:Q^{com}\times\Sigma_{com}\rightarrow 2^{Q^{com}}$ is defined as follows: For each $q\in Q^{com}$ and each $a \in \Sigma_{com}$, let $\mu_q:=\max_{(m_{\gamma},i)\in q}i$, and $q'\in\delta^{com}(q,a)$ if one of the following holds; otherwise, $\delta^{com}(q,a)=\varnothing$:
\begin{enumerate}
\item $a=\gamma^{in}$, $\mu_q\geq 1$ and 
\[q'=\{m_{\gamma'},i-1)\in S_{com}|(m_{\sigma'},i)\in q\}\cup \{(m_{\gamma},num^c)\}.\]
\item $a=\gamma^{out}$ and $q'\in \cup_{i:(m_{\gamma},i)\in q}\{q-\{(m_{\gamma},i)\}\}$.
\item $a=\gamma^{loss}$, $\sigma\in\Gamma_l$, and $q'\in \cup_{i:(m_{\gamma},i)\in q}\{q-\{(m_{\gamma},i)\}\}$.
\end{enumerate}
Intuitively, Rule
1 says that the sending of a control pattern $\gamma^{in}$ by the supervisor $S$ causes two consequences, i.e., the timer counts down by one unit for all messages in the channel and a new message $(m_{\gamma},num^c)$ is added into the channel. Rule 2 says that the firing of event $\gamma^{out}$ removes some message $(m_{\gamma},i)$ from the channel. Rule 3 says that some message $(m_{\gamma},i)$ with $\gamma\in\Gamma_l$ may get lost.

The output of the $CC$ model is a control pattern $\gamma^{out}\in\Gamma^{out}$, which cannot be recognized by the plant $G$, whose alphabet is $\Sigma^{in}\cup\Sigma_{uo}$. To link up these two models, we need to create an interface called {\em the command execution automaton} $G_{CE}$, which maps each control pattern $\gamma^{out}$ to a set of events in $\Sigma^{in}\cup\Sigma_{uo}$. Let $G_{CE}=(Q^{CE}, \Sigma_{CE}, \delta^{CE},q_0^{CE})$, where $Q^{CE}=\{q^{\gamma} \mid \gamma \in \Gamma\} \cup \{q_{wait}\}$, $\Sigma_{CE}=\Gamma^{out} \cup \Sigma_{o}^{in} \cup \Sigma_{uo}$, $q_0^{CE}=q_{wait}$. $\delta^{CE}: Q^{CE} \times \Sigma_{CE} \rightarrow Q^{CE}$ is defined as follows. 
 \begin{enumerate}
     \item for any $\sigma \in \Sigma_u \cap \Sigma_{uo}$, $\delta^{CE}(q_{wait}, \sigma)=q_{wait}$,
     \item for any $\sigma \in \Sigma_u \cap \Sigma_{o}$, $\delta^{CE}(q_{wait}, \sigma^{in})=q_{wait}$,
      \item for any $\gamma \in \Gamma$, $\delta^{CE}(q_{wait},\gamma^{out})=q^{\gamma}$,
     \item for any $\gamma, \gamma' \in \Gamma$, $\delta^{CE}(q^{\gamma}, \gamma'^{out})=q^{\gamma}$,
     \item for any $q^\gamma$, if $\sigma \in \Sigma_{o} \cap (\gamma \cup \Sigma_u)$, $\delta^{CE}(q^{\gamma},\sigma^{in})=q_{wait}$,
      \item for any $q^\gamma$, if $\sigma \in \Sigma_{uo} \cap (\gamma \cup \Sigma_u)$, $\delta^{CE}(q^{\gamma},\sigma)=q^{\gamma}$,
     \item and no other transitions are defined.
 \end{enumerate}

 Intuitively, at the initial state $q_{wait}$, $G_{CE}$ waits
to receive a control message, while in the mean time any uncontrollable event can be executed and will only lead to a self-loop at $q_{wait}$. This is reflected in Rules 1), 2). Rule 3) says that once a control message $m_{\gamma}$ is received, it transits to state $q^{\gamma}$ that records this most recently received control command, which will be used next. Any other control commands received within the same time step will be ignored, leading to a self-loop at state $q^{\gamma}$, which is reflected in Rule 4). Then, only those events in $\gamma \cup \Sigma_u$ are allowed to be fired. If an observable event $\sigma \in \Sigma_o \cap (\gamma \cup \Sigma_u)$ is fired at state $q^{\gamma}$, then $G_{CE}$ returns to the initial state $q_{wait}$, waiting to receive a new control command; if an unobservable event $\sigma \in \Sigma_{uo} \cap (\gamma \cup \Sigma_u)$ is fired at $q^{\gamma}$ instead, the command execution automaton self-loops at state $q^{\gamma}$ as the same control command $\gamma$ is to be used for the next event execution. This is reflected in Rule (5) and Rule (6), respectively. In particular, if an unobservable event $\sigma$ is fired at state $q^{\gamma}$, the control commands received within the next time step will be thrown away, as reflected in Rule (4) and Rule (6) combined. 

In the case that the original plant is $G=(Q, \Sigma, \delta, q_0, Q_m)$, we replace each $\sigma\in\Sigma_o$ with $\sigma^{in}\in\Sigma_o^{in}$ and create a new plant model $G^{mod} = (Q, \Sigma_{mod}=\Sigma_o^{in}\cup\Sigma_{uo}, \delta^{mod}, q_0, Q_m)$, where  for any $\sigma \in \Sigma_{uo}$, $\delta^{mod}(q,\sigma)=q'$ iff  $\delta(q,\sigma)=q'$, and for any $\sigma \in \Sigma_o$, $\delta^{mod}(q,\sigma^{in})=q'$ iff  $\delta(q,\sigma)=q'$. We now treat $G^{mod}$ as the plant, and rename it as $G$ in accordance with Fig.~\ref{fig:refined}. Thus, $G$ is over $\Sigma_{uo} \cup \Sigma_{o}^{in}$ (after relabelling). Let $\mathcal{P}:=G||G_{OC}||G_{CC}||G_{CE}$ be the new networked system plant, where the alphabet is $\Sigma^{\mathcal{P}}:=\Sigma_{uo}\cup\Sigma_o^{in}\cup\Sigma_o^{out}\cup\Sigma_o^{loss}\cup\Gamma^{in}\cup\Gamma^{out}\cup\Gamma^{loss}$, the controllable alphabet is $\Sigma_c^{\mathcal{P}}:=\Gamma^{in}\cup (\Sigma_{uo}\cap\Sigma_c)\cup \{\sigma^{in}\in\Sigma_o^{in}|\sigma\in\Sigma_c\}$, and the observable alphabet to the supervisor $S$ is $\Sigma_o^{\mathcal{P}}:=\Sigma_o^{out}\cup\Gamma^{in}$. We have the following networked control problem:
\begin{problem}
Given the networked plant $\mathcal{P}$ and a specification $E\subseteq L_m(G)$, design a supervisor $S$ over $\Sigma_o^{out}\cup\Sigma_o^{loss}\cup\Gamma^{in}$ such that
\begin{itemize}
\item $L_m(\mathcal{P}||S)\subseteq E$;
\item $\mathcal{P}||S$ is nonblocking;
\item $\mathcal{P}||S$ is {\em state-controllable} w.r.t. $\mathcal{P}$ \cite{su2010model} \cite{su2010aggregative};
\item $\mathcal{P}||S$ is {\em state-observable} w.r.t. $\mathcal{P}$ and $P_o$ \cite{su2010model} \cite{su2010aggregative}, where $P_o:(\Sigma^{\mathcal{P}})^*\rightarrow (\Sigma_o^{\mathcal{P}})^*$ is the natural projection.\hfill $\Box$
\end{itemize} 
\end{problem}
The problem can be solved by using automaton-based synthesis methods, e.g., \cite{su2010model} \cite{su2010aggregative} \cite{su2011synthesis}, which allows the plant $\mathcal{P}$ to be a nondeterministic finite-state automaton, and the final synthesized supervisor $S$ to be deterministic. The synthesis tool  \textbf{SUSYNA}  for solving supervisor synthesis for nondeterministic plants can be found at: https://www.ntu.edu.sg/home/rsu/Downloads.htm.  

In \cite{rashidinejad2018supervisory} the authors consider the setup where the (lossless) observation channel is non-FIFO while the (lossless) control channel is FIFO; both the channels are allowed to be of infinite capacities. The paper studies the networked supervisor synthesis problem for timed discrete-event systems; thus, the elapse of time is measured by the number of occurrences of ticks. Both the observation channel and the control channel are assumed to have a fixed communication delay, which may not be practical. Since activity loops are prohibited, the two channels are effectively reduced to be of  bounded capacities. To model the asynchronous interaction between the plant and the supervisor, the asynchronous product of the plant and the supervisor is used, which is equivalent to the standard synchronous product of the plant, the supervisor and the two channel models. To deal with observation delays and disorderings, an automaton is proposed which models the behaviour of the observed plant, i.e., the plant together with the observation channels. On the basis of this observed plant, a nonpredictive supervisor is synthesized that provides safety and nonblockingness for the observed plant, by slightly adapting the Bertil-Wonham framework of supervisor synthesis for timed discrete-event systems. To deal with control delays, the nonpredictive supervisor achieved for the observed plant is transformed to a networked supervisor that enables the events beforehand. To put their synthesis algorithm in the perspective of our general discussion, the nonpredictive supervisor synthesis step  is performed on the plant  $P'=OC\lVert CE \lVert G$ and transformed to the networked supervisor after taking the model of the control channel $CC$ into consideration. It is not known how such  a two-step approach can deal with setups with lossy channels, non-FIFO control channel, channels with bounded delays in a flexible manner. The assumption that  all plant events are observable may also be unrealistic. The asynchronous product operation defined in \cite{rashidinejad2018supervisory} is also presented in \cite{rashidinejad2019supervisory}, which considers the fact that enablement, execution, and observation of an event do not occur simultaneously but with some delay. 

\section{Discussions of existing challenges}
\label{sec:network_control_challenge}
After reviewing existing works on networked control of discrete-event systems, we can see that currently there are two main bodies of frameworks, based on channel delay models:
\begin{itemize}
\item F1 - an implicit channel delay model: in this framework, with \cite{lin2014control} being the representative work, the impact of channel delays on observability and controllability of a closed-loop system is explicitly envisioned, as captured by properly defined concepts of network observability and network controllability, without providing a detailed delay process model. 
\item F2 - an explicit channel delay mdoel: in this framework, with \cite{zhu2019supervisor} being the representative work, a detailed delay process model is explicitly given, upon which its impact on observability and controllability becomes part of the system analysis and control task, and is not explicitly embedded in those definitions.      
\end{itemize}
To illustrate the difference between these two frameworks intuitively, let $\mathcal{D}$ denotes the ($OC$ and/or $CC$) delay process, $\mathcal{C}(G,S,\mathcal{D})$ be the system controllability of $(G,S)$ under the influence of $\mathcal{D}$, and $\mathcal{O}(G,S,\mathcal{D})$ be the system observability of $(G,S)$ under the influence of $\mathcal{D}$.  The key research focus of networked control is to understand and precisely describe the following implications:
\begin{eqnarray}
\mathcal{D} &\Rightarrow & \mathcal{C}(G,S,\mathcal{D})\\
\mathcal{D} &\Rightarrow & \mathcal{O}(G,S,\mathcal{D})
\end{eqnarray}
In F1, because $\mathcal{D}$ is not precisely modelled, $\mathcal{C}(G,S,\mathcal{D})$ and $\mathcal{O}(G,S,\mathcal{D})$
need to be defined as brand new concepts. The catch is that it is unclear whether there is a specific physically realizable delay process $\mathcal{D}$ that make $\mathcal{C}(G,S,\mathcal{D})$ and $\mathcal{O}(G,S,\mathcal{D})$ physically feasible. In addition, the network controllability and observability concepts are typically very complicated, and hard to follow. In F2, by precisely modeling $\mathcal{D}$, the impact of $\mathcal{D}$ on the system can be precisely modelled as $G||\mathcal{D}$, which is then treated as a new plant. The concepts of network controllability and observability become $\mathcal{C}(G||\mathcal{D},S)$ and $\mathcal{O}(G||\mathcal{D},S)$, which are simply the standard concepts of controllability and observability in the classical supervisory control theory without explicitly mentioning delays. In other words, the actual impact of $\mathcal{D}$ becomes part of the plant behaviours. Thus, in principle, all existing synthesis methods such as centralized control, modular control, decentralized control, hiararchical control, and state-based control may be applied.   

Although channels may be either FIFO or non-FIFO in the literature, all existing networked control frameworks assume a target system $(G,S,\mathcal{D})$ to be asynchronous, which may not be applicable in reality, as communication channels typically operates in a concurrent manner, i.e., the message input and output of each single channel typically take place concurrently. It is unclear how concurrency can be handled in F1. But it could be handled in F2. For example, by considering an elaborated channel delay model based on the one proposed in \cite{zhu2019supervisor}, where events in $\Sigma_o^{in}$ and $\Sigma_o^{out}$ may take place either synchronously or asynchronously in the $OC$, and so do control messages in $\Gamma^{in}$ and $\Gamma^{out}$ in the $CC$. This essentially calls for a concurrent networked supervisory control framework, which shall match reality better.     

Computational complexity is always one major concern for supervisory control theory, which seems an even more daunting challenge for F1. How to efficient determine the existence of a networked supervisor and, in case it exists, how to efficiently compute it are one important problem to be solved. It is interesting to see whether we could borrow ideas from the minimal communication works proposed in \cite{ricker1999incorporating} \cite{ricker2008asymptotic} to handle both observation messages and control messages. Minimal communication is also important to enhance attack-resilience of networked systems, which shall continue to be one important research direction. Finally, it could be the time to consider a new supervisory control architecture, especially the supervisory control map, which might be more robust to channel delays than the standard Ramadge-Wonham supervisory control architecture. 

%
%

\bibliographystyle{unsrt}      
\bibliography{sample}   

\begin{thebibliography}{10}

\bibitem{yoo2002general}
T-S Yoo and St{\'e}phane Lafortune.
\newblock A general architecture for decentralized supervisory control of
  discrete-event systems.
\newblock {\em Discrete Event Dynamic Systems}, 12(3):335--377, 2002.

\bibitem{barrett2000decentralized}
George Barrett and St{\'e}phane Lafortune.
\newblock Decentralized supervisory control with communicating controllers.
\newblock {\em IEEE Transactions on Automatic Control}, 45(9):1620--1638, 2000.

\bibitem{ricker1999incorporating}
S~Laurie Ricker and Karen Rudie.
\newblock Incorporating communication and knowledge into decentralized
  discrete-event systems.
\newblock In {\em Proceedings of the 38th IEEE Conference on Decision and
  Control (Cat. No. 99CH36304)}, volume~2, pages 1326--1332. IEEE, 1999.

\bibitem{ricker2008asymptotic}
SL~Ricker.
\newblock Asymptotic minimal communication for decentralized discrete-event
  control.
\newblock In {\em 2008 9th International Workshop on Discrete Event Systems},
  pages 486--491. IEEE, 2008.

\bibitem{wang2008minimization}
Weilin Wang, St{\'e}phane Lafortune, and Feng Lin.
\newblock Minimization of communication of event occurrences in acyclic
  discrete event systems.
\newblock {\em IEEE Transactions on Automatic Control}, 53(9):2197--2202, 2008.

\bibitem{ricker2013overview}
Laurie Ricker.
\newblock An overview of synchronous communication for control of decentralized
  discrete-event systems.
\newblock In {\em Control of Discrete-Event Systems}, pages 127--146. Springer,
  2013.

\bibitem{tripakis2004decentralized}
S.~Tripakis.
\newblock Decentralized control of discrete-event systems with bounded or
  unbounded delay communication.
\newblock {\em IEEE Transactions on Automatic Control}, 49(9):1489--1501, 2004.

\bibitem{feng2008supervisory}
Lei Feng and Walter~Murray Wonham.
\newblock Supervisory control architecture for discrete-event systems.
\newblock {\em IEEE Transactions on Automatic Control}, 53(6):1449--1461, 2008.

\bibitem{komenda2015coordination}
Jan Komenda, Tom{\'a}{\v{s}} Masopust, and Jan~H van Schuppen.
\newblock Coordination control of discrete-event systems revisited.
\newblock {\em Discrete Event Dynamic Systems}, 25(1-2):65--94, 2015.

\bibitem{su2010model}
Rong Su, Jan~H van Schuppen, and Jacobus~E Rooda.
\newblock Model abstraction of nondeterministic finite-state automata in
  supervisor synthesis.
\newblock {\em IEEE Transactions on automatic control}, 55(11):2527--2541,
  2010.

\bibitem{su2010aggregative}
Rong Su, Jan~H Van~Schuppen, and Jacobus~E Rooda.
\newblock Aggregative synthesis of distributed supervisors based on automaton
  abstraction.
\newblock {\em IEEE Transactions on Automatic Control}, 55(7):1627--1640, 2010.

\bibitem{su2011synthesis}
Rong Su, Jan~H Van~Schuppen, and Jacobus~E Rooda.
\newblock The synthesis of time optimal supervisors by using heaps-of-pieces.
\newblock {\em IEEE Transactions on Automatic Control}, 57(1):105--118, 2011.

\bibitem{gaubert1995performance}
St{\'e}phane Gaubert.
\newblock Performance evaluation of (max,+) automata.
\newblock {\em IEEE transactions on automatic Control}, 40(12):2014--2025,
  1995.

\bibitem{park2007decentralized}
Seong-Jin Park and Kwang-Hyun Cho.
\newblock Decentralized supervisory control of discrete event systems with
  communication delays based on conjunctive and permissive decision structures.
\newblock {\em Automatica}, 43(4):738--743, 2007.

\bibitem{hiraishi2009solvability}
Kunihiko Hiraishi.
\newblock On solvability of a decentralized supervisory control problem with
  communication.
\newblock {\em IEEE Transactions on Automatic Control}, 54(3):468--480, 2009.

\bibitem{sadid2015robustness}
Waselul~Haque Sadid, Laurie Ricker, and Shahin Hashtrudi-Zad.
\newblock Robustness of synchronous communication protocols with delay for
  decentralized discrete-event control.
\newblock {\em Discrete Event Dynamic Systems}, 25(1-2):159--176, 2015.

\bibitem{zhang2016distributed}
R.~Zhang, K.~Cai, Y.~Gan, and W.M. Wonham.
\newblock Distributed supervisory control of discrete-event systems with
  communication delay.
\newblock {\em Discrete Event Dynamic Systems}, 26(2):263--293, 2016.

\bibitem{zhang2016delay}
Renyuan Zhang, Kai Cai, Yongmei Gan, and WM~Wonham.
\newblock Delay-robustness in distributed control of timed discrete-event
  systems based on supervisor localisation.
\newblock {\em International Journal of Control}, 89(10):2055--2072, 2016.

\bibitem{kalyon2011synthesis}
Gabriel Kalyon, Tristan Le~Gall, Herv{\'e} Marchand, and Thierry Massart.
\newblock Synthesis of communicating controllers for distributed systems.
\newblock In {\em 2011 50th IEEE Conference on Decision and Control and
  European Control Conference}, pages 1803--1810. IEEE, 2011.

\bibitem{kalyon2013symbolic}
Gabriel Kalyon, Tristan Le~Gall, Herv Marchand, and Thierry Massart.
\newblock Symbolic supervisory control of distributed systems with
  communications.
\newblock {\em IEEE Transactions on Automatic Control}, 59(2):396--408, 2013.

\bibitem{darondeau2012distributed}
Philippe Darondeau and Laurie Ricker.
\newblock Distributed control of discrete-event systems: A first step.
\newblock In {\em Transactions on Petri Nets and Other Models of Concurrency
  VI}, pages 24--45. Springer, 2012.

\bibitem{alves2017supervisory}
M.~V.~S. Alves, L.~K. Carvalho, and J.~C. Basilio.
\newblock Supervisory control of timed networked discrete event systems.
\newblock {\em Conference on Decision and Control}, 56:4859--4865, 2017.

\bibitem{alves2019robust}
Marcos~VS Alves, Antonio~EC da~Cunha, Lilian~Kawakami Carvalho, Marcos~Vicente
  Moreira, and Jo{\~a}o~Carlos Basilio.
\newblock Robust supervisory control of discrete event systems against
  intermittent loss of observations.
\newblock {\em International Journal of Control}, pages 1--13, 2019.

\bibitem{zhou2019supervisory}
Lei Zhou, Shaolong Shu, and Feng Lin.
\newblock Supervisory control of discrete event systems under nondeterministic
  observations.
\newblock In {\em 2019 18th European Control Conference (ECC)}, pages
  4192--4197. IEEE, 2019.

\bibitem{sasi2018detectability}
Yazeed Sasi and Feng Lin.
\newblock Detectability of networked discrete event systems.
\newblock {\em Discrete Event Dynamic Systems}, 28(3):449--470, 2018.

\bibitem{lin2019state}
Feng Lin, Weilin Wang, Leitao Han, and Bin Shen.
\newblock State estimation of multichannel networked discrete event systems.
\newblock {\em IEEE Transactions on Control of Network Systems}, 7(1):53--63,
  2019.

\bibitem{alves2019state}
Marcos~VS Alves and Jo{\~a}o~C Basilio.
\newblock State estimation and detectability of networked discrete event
  systems with multi-channel communication networks.
\newblock In {\em 2019 American Control Conference (ACC)}, pages 5602--5607.
  IEEE, 2019.

\bibitem{balemi1992supervision}
S~Balemi and UA~Brunner.
\newblock Supervision of discrete event systems with communication delays.
\newblock In {\em 1992 American Control Conference}, pages 2794--2798. IEEE,
  1992.

\bibitem{balemi1994input}
S.~Balemi.
\newblock Input/output discrete event processes and communication delays.
\newblock {\em Discrete Event Dynamic Systems}, 4(1):41--85, 1994.

\bibitem{park2006delay}
Seong-Jin Park and Kwang-Hyun Cho.
\newblock Delay-robust supervisory control of discrete-event systems with
  bounded communication delays.
\newblock {\em IEEE Transactions on Automatic Control}, 51(5):911--915, 2006.

\bibitem{park2007supervisory}
Seong-Jin Park and Kwang-Hyun Cho.
\newblock Supervisory control of discrete event systems with communication
  delays and partial observations.
\newblock {\em Systems \& control letters}, 56(2):106--112, 2007.

\bibitem{lin2014control}
F.~Lin.
\newblock Control of networked discrete event systems: dealing with
  communication delays and losses.
\newblock {\em SIAM Journal on Control and Optimization}, 52(2):1276--1298,
  2014.

\bibitem{shu2014supervisor}
Shaolong Shu and Feng Lin.
\newblock Supervisor synthesis for networked discrete event systems with
  communication delays.
\newblock {\em IEEE Transactions on Automatic Control}, 60(8):2183--2188, 2014.

\bibitem{shu2016deterministic}
Shaolong Shu and Feng Lin.
\newblock Deterministic networked control of discrete event systems with
  nondeterministic communication delays.
\newblock {\em IEEE Transactions on Automatic Control}, 62(1):190--205, 2016.

\bibitem{wang2016robust}
F.~Wang, S.~Shu, and F.~Lin.
\newblock Robust networked control of discrete event systems.
\newblock {\em IEEE Transactions on Automation Science and Engineering},
  13(4):1528--1540, 2016.

\bibitem{shu2016predictive}
S.~Shu and F.~Lin.
\newblock Predictive networked control of discrete event systems.
\newblock {\em IEEE Transactions on Automatic Control}, 62(9):4698--4705, 2016.

\bibitem{zhao2015supervisory}
Bo~Zhao, Feng Lin, Caisheng Wang, Xuesong Zhang, Michael~P Polis, et~al.
\newblock Supervisory control of networked timed discrete event systems and its
  applications to power distribution networks.
\newblock {\em IEEE Transactions on Control of Network Systems}, 4(2):146--158,
  2015.

\bibitem{hou2019relative}
Yunfeng Hou, Weilin Wang, Yanwei Zang, Feng Lin, Miao Yu, and Chaohui Gong.
\newblock Relative network observability and its relation with network
  observability.
\newblock {\em IEEE Transactions on Automatic Control}, 2019.

\bibitem{shu2014decentralized}
S.~Shu and F.~Lin.
\newblock Decentralized control of networked discrete event systems with
  communication delays.
\newblock {\em Automatica}, 50(8):2108--2112, 2014.

\bibitem{komenda2016modular}
J.~Komenda and F.~Lin.
\newblock Modular supervisory control of networked discrete-event systems.
\newblock In {\em 2016 13th International Workshop on Discrete Event Systems
  (WODES)}, pages 85--90. IEEE, 2016.

\bibitem{komenda2015distributed}
Jan Komenda, Tom{\'a}{\v{s}} Masopust, and Jan~H van Schuppen.
\newblock On a distributed computation of supervisors in modular supervisory
  control.
\newblock In {\em 2015 International Conference on Complex Systems Engineering
  (ICCSE)}, pages 1--6. IEEE, 2015.

\bibitem{wong2004computation}
Kai~C Wong and Walter~Murray Wonham.
\newblock On the computation of observers in discrete-event systems.
\newblock {\em Discrete Event Dynamic Systems}, 14(1):55--107, 2004.

\bibitem{zhu2019supervisor}
Yuting Zhu, Liyong Lin, Simon Ware, and Rong Su.
\newblock Supervisor synthesis for networked discrete event systems with
  communication delays and lossy channels.
\newblock In {\em 2019 IEEE 58th Conference on Decision and Control (CDC)},
  pages 6730--6735. IEEE, 2019.

\bibitem{rashidinejad2018supervisory}
Aida Rashidinejad, Michel Reniers, and Lei Feng.
\newblock Supervisory control of timed discrete-event systems subject to
  communication delays and non-fifo observations.
\newblock {\em IFAC-PapersOnLine}, 51(7):456--463, 2018.

\bibitem{rashidinejad2019supervisory}
Aida Rashidinejad, Michel Reniers, and Martin Fabian.
\newblock Supervisory control of discrete-event systems in an asynchronous
  setting.
\newblock In {\em 2019 IEEE 15th International Conference on Automation Science
  and Engineering (CASE)}, pages 494--501. IEEE, 2019.

\end{thebibliography}

\end{document}